\documentstyle[12pt,amstex,righttag]{article}
\input epsf.tex

\newcommand{\bra}{\langle}
\newcommand{\ket}{\rangle}
\newcommand{\ba}{\begin{matrix}}
\newcommand{\ea}{\end{matrix}}
\newcommand{\CR}{\nonumber \\}

\newcommand{\SW}{\lambda_{\text{SW}}}
\newcommand{\bo}[1]{\boldsymbol{#1}}
\newcommand{\cN}{{\cal N}}
\newcommand{\wt}{\widetilde}
\newcommand{\wh}{\widehat}
\newcommand{\coh}{coherent\ }
\newcommand{\K}{K\"ahler\ }
\newcommand{\cy}{Calabi--Yau\ }
\newcommand{\dP}{del Pezzo\ }

\newcommand{\PF}{Picard--Fuchs\ }
\newcommand{\Todd}{{\text{Todd}}}          % Todd class
\newcommand{\ch}{{\text{ch}}}              % Chern character
\newcommand{\cE}{{\cal E}}              % sheaf
\newcommand{\cG}{{\cal G}}              % sheaf
\newcommand{\cO}{{\cal O}}              % structure sheaf
\newcommand{\BN}{{B}_N}                   % Nth del Pezzo surface
\newcommand{\Bnine}{{B}_9}                   % 9th del Pezzo surface
\newcommand{\EN}{\widehat{E}_N}              % affine E_N
\newcommand{\bZ}{\bo{Z}}

\newcommand{\bP}{\bo{P}}
\newcommand{\om}{\bo{\omega}}
\newcommand{\cfS}{c_1(S)}           % 1st Chern of S
\newcommand{\csS}{c_2(S)}           % 2nd Chern of S
\newcommand{\cfBN}{c_1(\BN)}        % 1st Chern of B_N
\newcommand{\csBN}{c_2(\BN)}        % 2nd Chern of B_N
\newcommand{\scirc}{\text{\small $\circ$}}
\begin{document}
\begin{titlepage}
\begin{flushright}
{\tt hep-th/0007243} \\
UTHEP-431 \\
July, 2000
\end{flushright}
\vspace{0.5cm}
\begin{center}
{\Large \bf 
Duality Between String Junctions and D-Branes on Del Pezzo Surfaces
\par}
\lineskip .75em
\vskip2.5cm
{\large Kenji Mohri, Yukiko Ohtake and Sung-Kil Yang}
\vskip 1.5em
{\large\it Institute of Physics, University of Tsukuba \\
\vspace{1mm}
Ibaraki 305-8571, Japan}
\end{center}
\vskip3cm
\begin{abstract}
We revisit local mirror symmetry associated with del Pezzo surfaces in
\cy threefolds in view of five-dimensional $\cN$=1 $E_N$ theories
compactified on a circle. The mirror partner of singular \cy with
a shrinking del Pezzo four-cycle is described as the affine
7-brane backgrounds probed by a D3-brane. Evaluating the mirror map and the 
BPS central charge we relate junction charges to RR charges of D-branes wrapped
on del Pezzo surfaces. This enables us to determine how the string junctions
are mapped to D-branes on del Pezzo surfaces.
\end{abstract}
\end{titlepage}
\baselineskip=0.7cm

%%%%%%%%%%%%%%%%%%%%%%
\section{Introduction}
%%%%%%%%%%%%%%%%%%%%%%
Five-dimensional (5D) $\cN$=1 theories with global exceptional symmetries
are non-trivial interacting superconformal theories \cite{S}. They appear 
in the strong-coupling limit of 5D $\cN$=1 supersymmetric $SU(2)$ gauge theory
with $N_f$ quark hypermultiplets. For $N_f \leq 8$ the microscopic global 
symmetry $SO(2N_f)\times U(1)$ is enhanced to $E_{N_f+1}$. When $N_f=0$
we have two theories with global $E_1$ and $\wt E_1$ symmetry. The $\wt E_1$ 
theory is shown to further flow down to the $\wt E_0$ theory which has no 
global symmetry.

It is well-known that these 5D $E_N$ theories are obtained by
compactifying M-theory on a \cy threefold with a shrinking
four-cycle realized by del Pezzo surfaces \cite{MS,DKV}. 
Then the geometrical meaning of the sequence of $E$-type symmetries 
is most naturally understood in terms of 
the blowing-up process of $\bo{P}^2$ (or of $\bo{P}^1\times\bo{P}^1$) which 
yields del Pezzo surfaces.
The correspondence is summarized in Table 1.

\renewcommand{\arraystretch}{1.7}
\begin{table}[b]
\begin{center}
\begin{tabular}{||c|c|c||} \hline
$\wt E_0$   &   $E_1$  &  $\wt E_1$, $E_N$ $(2 \leq N\leq 8)$ \\
\hline
$\bo{P}^2$   &   $\bo{P}^1 \times \bo{P}^1$  &  $\bo{P}^2$ with the $N$ points blown up \\
\hline
\end{tabular}
\end{center}
\caption{$E$-type symmetries and del Pezzo surfaces}
\label{tbl1}
\end{table}
\renewcommand{\arraystretch}{1}

Another interesting realization of 5D $E$-type theories is provided by the IIB
5-brane web construction including background 7-branes \cite{DHIK}. 
The advantageous point in 
this is that the affine property of $E_N$ symmetry, which has been known to
occur naturally as the action of the Weyl group of the affine algebra $\wh E_N$
in the study of the del Pezzo surfaces \cite{L}, can be captured explicitly
by the 7-brane configurations, thanks to the recent advances in 7-brane
technology \cite{D,DHIZ}.

Furthermore it is shown that the Coulomb branch of 5D $E_N$ theories 
compactified on a circle is described in terms of a D3-brane probing the
affine 7-brane backgrounds \cite{YY}. This provides us with an intuitive 
physical picture behind important calculations performed in \cite{LMW, MNW}. 
Then, in the light of the analysis of \cite{LMW} and the 7-brane picture, 
we see that the $E_N$ theories on $\bo{R}^4 \times S^1$ is considered as 
either the D3--7-brane system or the local system realized in singular 
\cy space with a shrinking del Pezzo four-cycle; these two systems are 
mirror dual to each other. 
In this paper, thus, we revisit local mirror symmetry associated 
with del Pezzo surfaces in \cy threefolds and elucidate on the
relation between IIB string junctions in the affine 7-brane backgrounds and
IIA D-branes wrapped on del Pezzo surfaces. A map between these two objects 
has been worked out recently in \cite{HI} in which an isomorphism between the 
junction lattices and the even homology lattice of del Pezzo surfaces,
which is identified with the Ramond-Ramond (RR) charge lattice of IIA 
D-branes, is shown by comparing the intersection forms. 
In our approach, on the other hand, we evaluate explicitly the mirror 
map and the BPS central charge so as to convert junction charges 
to the central charges of D-branes.

The paper is organized as follows: In the next section, we calculate in detail
the Seiberg--Witten (SW) period integrals which describe the Coulomb branch of
5D $E_N$ theories on $\bo{R}^4 \times S^1$. The monodromy matrices and the
prepotentials are obtained. In section 3, the mirror map is constructed.
In section 4, we analyze  D-branes localized on a surface.
Several important invariants of the BPS D-branes such as
the RR charge, the central charge and intersection pairings are
given in algebro-geometrical terms.  
In section 5, the results of the preceding section are utilized to 
verify the map between string junctions and D-branes on a
del Pezzo four-cycle. 

%%%%%%%%%%%%%%%%
\section{Calculation of the periods and monodromies}
%%%%%%%%%%%%%%%%
\renewcommand{\theequation}{2.\arabic{equation}}\setcounter{equation}{0}

The elliptic curves for the $\wh{E}_N$ 7-brane configurations are 
obtained in \cite{YY,SZ}. 
These curves describe the Coulomb branch of 5D $E_N$ theories compactified 
on $S^1$ \cite{YY}. For $E_{N=8,7,6}$ theories, they are given 
by \cite{YY,LMW,MNW}
\begin{align}
\wh{E}_8 &\quad:\quad 
y^2 = x^3 + R^2 u^2 x^2 -2 u^5, \CR
\wh{E}_7 &\quad:\quad 
y^2 = x^3 + R^2 u^2 x^2 +2 u^3 x, \label{eqn:SW}\\
\wh{E}_6 &\quad:\quad
y^2 = x^3 + R^2 u^2 x^2 -2 R i u^3 x -u^4,\nonumber
\end{align}
where $u$ is a complex moduli parameter with mass dimension 6, 4, 3 for 
$\wh{E}_8$, $\wh{E}_7$, $\wh{E}_6$ and $R$ is the radius of $S^1$. 
These curves have the discriminant
\begin{align}
\wh{E}_8 &\quad:\quad \Delta = -4 u^{10} (2 R^6 u -27),\CR
\wh{E}_7 &\quad:\quad \Delta = -4 u^9 (R^4 u -8),\\
\wh{E}_6 &\quad:\quad \Delta = -i u^8 (4 R^3 u +27 i),\nonumber
\end{align}
whose zeroes at $u=0$ represent coalescing $(N+2)$ 7-branes of $E_{N=8,7,6}$
configurations and a single zero at $u\neq 0$ represents a 7-brane which is 
responsible for extending $E_N$ to the affine system.
The SW differentials $\SW$ associated with the $\wh{E}_N$ curves are 
defined by
\begin{equation}
\frac{d \SW}{du}=\frac{dx}{y}+d(*),
\end{equation}
which are known to take the logarithmic form \cite{LMW,MNW}. 
For the curves (\ref{eqn:SW}) we find
\begin{align}
\wh{E}_8 &\quad:\quad \SW = 
\frac{\kappa}{2\pi iR}\log\left(\frac{y-Rux}{y+Rux}\right)\frac{dx}{x},\CR
\wh{E}_7 &\quad:\quad \SW = 
\frac{\kappa}{2\pi iR}\log\left(\frac{y-Rux}{y+Rux}\right)\frac{dx}{x},\\
\wh{E}_6 &\quad:\quad \SW =
\frac{\kappa}{2\pi iR}\log\left(\frac{y-Rux+iu^2}{y+Rux-iu^2}\right)\frac{dx}{x-iu/R},
\nonumber
\end{align}
where $\kappa$ is a normalization constant and the factor $1/R$ ensures that 
$\SW$ has mass dimension unity.
Notice that $\SW$ possesses the pole at $x=0$ for $\wh{E}_8$, $\wh{E}_7$
and $x=iu/R$ for $\wh{E}_6$
because of the multivaluedness of the logarithm. 
Hence the set of period integrals $\Pi=( s, a, a_D )$ consists of 
\begin{align}
s&=\oint_{C}\SW =\frac{2\pi i\kappa}{R},\label{eqn:period1}\\
a(u)&=\oint_A\SW =\int^u du'\, \varpi(u'),\label{eqn:period2}\\
a_D(u)&=\oint_B\SW =\int^u du'\, \varpi_D(u'),\label{eqn:period3}
\end{align}
where the contour $C$ surrounds the pole of $\SW$ and 
$A$, $B$ are the homology cycles on $\wh{E}_N$ torus along which the torus 
periods $\varpi(u)$, $\varpi_D(u)$ are defined as usual,
\begin{equation}
\varpi(u)=\oint_A \frac{dx}{y}, \quad
\varpi_D(u)=\oint_B \frac{dx}{y}. 
\end{equation}

\subsection{\PF equations}

The period $\Pi$ is evaluated with the use of the \PF equations. 
It is convenient to introduce a dimensionless variable 
\begin{equation}
z = \frac{2}{27} R^6 u,\quad
\frac{1}{8} R^4 u,\quad
\frac{4i}{27} R^3 u
\end{equation}
for $\wh{E}_8$, $\wh{E}_7$, $\wh{E}_6$, respectively, 
to write down the \PF equations. 
We obtain
\begin{equation}
{\cal L}_{\wh{E}_N} \Pi(z) = 0, 
\end{equation}
where
\begin{align}
{\cal L}_{\wh{E}_8}&= 
\left[ 36 z^2 (z-1) \frac{d^2}{dz^2}
+ 4 z (27 z -18 ) \frac{d}{dz} 
+ (36 z - 5)\right] \frac{d}{dz}, \CR
{\cal L}_{\wh{E}_7}&= 
\left[ 16 z^2 (z-1) \frac{d^2}{dz^2} 
+ 16 z(3z-2)\frac{d}{dz} 
+(16z-3)\right] \frac{d}{dz}, 
\label{eqn:PF1}\\
{\cal L}_{\wh{E}_6}&= 
\left[ 9 z^2 (z-1) \frac{d^2}{dz^2} 
+z(27 z-18) \frac{d}{dz} 
+(9z-2)\right] \frac{d}{dz}. \nonumber
\end{align}
In order to derive the solution we first solve the second-order equation for 
$d\Pi /dz$, and then integrate over $z$ to get $\Pi(z)$.
Substituting the form $d\Pi /dz = z^{\rho} F(z)$ it is seen that
$\rho=-\frac{5}{6}$, $-\frac{3}{4}$, $-\frac{2}{3}$ for $\wh{E}_8$, 
$\wh{E}_7$, $\wh{E}_6$,
and $F(z)$ obeys the standard hypergeometric equation
\begin{equation}
\left[z(1-z)\frac{d^2}{dz^2}
+\left(\gamma-(\alpha+\beta+1)z\right)\frac{d}{dz}
-\alpha\beta\right]
F(z)=0,
\end{equation}
with $\alpha=\beta=\rho+1$ and $\gamma=2(\rho+1)$, that is,
\begin{equation}
\alpha=\frac{1}{6},\quad \frac{1}{4},\quad \frac{1}{3}
\end{equation}
for $\wh{E}_8, \wh{E}_7, \wh{E}_6$.\footnote[2]{We shall use the 
notation $\alpha$ specifically to denote these numbers throughout 
this paper.} 
The torus periods around the regular singular points at $z=0, 1, 
\infty$ are then expressed as
\begin{equation}
\begin{pmatrix} 
\varpi_D^{(\bullet)}(z) \\ 
\varpi^{(\bullet)}(z) 
\end{pmatrix}
= C_{\bullet}
\begin{pmatrix}
\varphi_1^{(\bullet)}(z) \\ 
\varphi_2^{(\bullet)}(z) 
\end{pmatrix},
\end{equation}
where $\bullet$ stands for $ 0,1, \infty$ corresponding to $z = 0,1,\infty$, 
$C_{\bullet}$ are $2\times 2$ coefficient matrices and the set of fundamental 
solutions has been taken as follows:
\begin{equation}
\begin{cases}
\varphi^{(0)}_1(z) =  
z^{-(1-\alpha)} {}_2F_1(\alpha, \alpha \,; 2\alpha \,; z),\\
 \varphi^{(0)}_2(z) =  z^{-\alpha} {}_2F_1(1\!-\!\alpha, 1\!-\!\alpha \,; 
2(1\!-\!\alpha)\,;\,z),
\end{cases}
\label{eqn:bases}
\end{equation}
\begin{equation}
\begin{cases}
 \varphi^{(1)}_1(z) = z^{-(1-\alpha)} {}_2F_1(\alpha, \alpha \,; 1\,; 1\!-\!z),\\
 \varphi^{(1)}_2(z) = z^{-(1-\alpha)} 
	\left({}_2F_1(\alpha, \alpha, \,; 1 \,; 1\!-\!z) \log(1\!-\!z) 
	+ {{}_2F_1}^*(\alpha, \alpha \,; 1 \,; 1\!-\!z)\right),
\end{cases}
\end{equation}
and 
\begin{equation}
\begin{cases}
 \varphi^{(\infty)}_1(z) = \frac{1}{z} \,
	{}_2F_1(\alpha, 1\!-\!\alpha\,; 1 \,; \frac{1}{z}),\\
 \varphi^{(\infty)}_2(z) = \frac{1}{z} 
	\left({}_2F_1(\alpha, 1\!-\!\alpha\,; 1 \,; \frac{1}{z}) \log(-z)
	+ {{}_2F_1}^*(\alpha, 1\!-\!\alpha\,; 1 \,; \frac{1}{z})\right).
\end{cases}
\end{equation}
Here ${}_2F_1(\alpha, \beta\,; \gamma\,; z)$ is the hypergeometric function 
\begin{equation}
{}_2F_1(\alpha, \beta; \gamma; z)= 
 \sum_{n=0}^{\infty} \frac{(\alpha)_n(\beta)_n}{(\gamma)_n}\frac{z^n}{n!}
\end{equation}
and
\begin{equation}
{{}_2F_1}^*(\alpha, \beta; \gamma; z)
=\sum_{n=1}^{\infty} \frac{(\alpha)_n(\beta)_n}{(\gamma)_n}
\left[ \sum_{k=0}^{n-1}\left(\frac{1}{\alpha+k}+\frac{1}{\beta+k}-\frac{
2}{\gamma+k}\right)\right]
\frac{z^n}{n!},
\end{equation}
where $(\alpha)_n=\Gamma(\alpha+n)/\Gamma(\alpha)$ with $\Gamma(x)$ being 
the gamma function.

The coefficient matrix $C_1$ is determined by directly evaluating the elliptic 
integrals. The result reads
\begin{equation}
C_1=\tilde{c}\,R^{\frac{1-\alpha}{\alpha}}
\left(\ba
-\pi				     & 0  \\
\frac{i}{2}\left(\log v-i\pi \right) & -\frac{i}{2}
\ea \right)
\end{equation}
where $\tilde{c}=2^2/3^3$, $2^{-3/2}$, $2^{5/2}/3^2\,$\footnote[3]{In what 
follows we will set $\tilde{c}=1$ for simplicity.} and
\begin{equation}
v=432, \quad 64, \quad 27
\end{equation}
for $\wh{E}_8, \wh{E}_7, \wh{E}_6$. 
Then, performing the analytic continuation we calculate $C_0$ and 
$C_{\infty}$. 
The connection matrices $X, Y$ defined by 
\begin{equation}
\left(\ba
\varphi^{(0)}_1 \\ \varphi^{(0)}_2
\ea \right)
=X \left(\ba
\varphi^{(1)}_1 \\ \varphi^{(1)}_2
\ea \right),
\quad
\left(\ba
\varphi^{(0)}_1 \\ \varphi^{(0)}_2
\ea \right)
=Y \left(\ba
\varphi^{(\infty)}_1 \\ \varphi^{(\infty)}_2
\ea \right)
\end{equation}
are found with the aid of the Barnes' integral representation of the 
hypergeometric function \cite[p.~286]{WW}.
It turns out that
\begin{equation}
X=\left(\ba
\xi_1\eta_1 & -\xi_1\\
\xi_2\eta_2 & -\xi_2
\ea\right), \quad
Y=\left(\ba
\xi_1\eta_1 \,e^{i\pi\alpha} & \xi_1 \,e^{i\pi\alpha}\\
\xi_2\eta_2 \,e^{i\pi(1-\alpha)} & \xi_2 \,e^{i\pi(1-\alpha)}
\ea\right),
\end{equation}
where 
\begin{align}
\xi_1 = \frac{\Gamma(2\alpha)}{\Gamma^2(\alpha)},  &\quad
\xi_2=\frac{\Gamma(2-2\alpha)}{\Gamma^2(1-\alpha)} ,\\
\eta_1 = 2(\psi(1)-\psi(\alpha)),  &\quad
\eta_2=2(\psi(1)-\psi(1-\alpha)), 
\end{align}
and $\psi(x)=\frac{d}{dx}\log\Gamma(x)$ is the digamma function. 
Thus we obtain 
\begin{align}
C_0 &= C_1 X^{-1} = \frac{\pi R^{1-\alpha \over \alpha}}{2(2\alpha-1)}
\left(\ba
2\xi_2& -2\xi_1\\
-\frac{\omega}{\sin \pi\alpha}\xi_2& \frac{\bar{\omega}}{\sin\pi\alpha}\xi_1
\ea \right),
\quad
\omega=e^{i\pi\left(\frac{1}{2}-\alpha\right)},\\
C_{\infty} &= C_1 X^{-1} Y 
= R^{1-\alpha \over \alpha}\left(\ba
-(\log v+i\pi)\sin\pi\alpha&-\sin\pi\alpha\\
\frac{i\pi}{2\sin\pi\alpha}&0
\ea\right).
\end{align}

When the periods go around each regular singular point counterclockwise 
they undergo the monodromy. 
If we denote as $T_{\bullet}$ the monodromy matrix of 
$\left(\ba\varphi_1^{(\bullet)}\\\varphi_2^{(\bullet)}\ea\right)$,
the monodromy matrix $\hat{M}_{\bullet}$ acting on 
$\left(\ba\varpi_D^{(\bullet)}\\ \varpi^{(\bullet)}\ea\right)$ is given
by $\hat{M}_{\bullet}=C_{\bullet}T_{\bullet}C_{\bullet}^{-1}$. 
For $\wh{E}_{N=8,7,6}$ one computes
\begin{equation}
\hat{M}_0 =\left(\ba
N-8& N-9\\
1 & 1
\ea\right),\quad
\hat{M}_1 =\left(\ba
1& 0\\
-1 & 1
\ea\right),\quad
\hat{M}_{\infty} =\left(\ba
1& N-9\\
0 & 1
\ea\right)
\label{eqn:monodromy}
\end{equation}
which obey $\hat{M}_{\infty}=\hat{M}_0\hat{M}_1$, $\hat{M}_0^6={1}$
($\wh{E}_8$), $\hat{M}_0^4={1}$ ($\wh{E}_7$), $\hat{M}_0^3={1}$
($\wh{E}_6$).
If we follow the convention in \cite{DHIZ}, 
$K_{\bullet}=\hat{M}_{\bullet}^{-1}$
yields the monodromy around the 7-brane configurations.
For the $[p,q]$ 7-brane $\bo{X}_{[p,q]}$, the monodromy matrix 
reads \cite{DHIZ}
\begin{equation}
K_{[p,q]}=\left(\ba
1+pq & -p^2\\
q^2 & 1-pq
\ea\right),
\end{equation}
whereas for the $\wh{E}_N$ 7-branes we have
\begin{equation}
K(\bo{\hat{E}}_N)=\left(\ba
1& 9-N\\
0& 1
\ea\right).
\end{equation}
In (\ref{eqn:monodromy}) we observe $\hat{M}_{\infty}^{-1}=K(\bo{\hat{E}}_N)$,
$\hat{M}_1^{-1}=K_{[0,1]}$ and $\hat{M}_0^{-1}=K_{[3,-2]}K_{[3,-1]}K_{[1,0]}^N$.
Accordingly our 7-brane configuration is identified as 
\begin{equation}
\bo{\hat{E}}_N=\bo{A}^N \bo{X}_{[3,-1]}\bo{X}_{[3,-2]}
\bo{X}_{[0,1]},
\end{equation}
where we have used the notation in \cite{DHIZ} and $\bo{A}=\bo{X}_{[1,0]}$ 
is the D7-brane.
This is shown to be equivalent to the canonical one $\bo{\hat{E}}_N=
\bo{A}^{N-1}\bo{B}\bo{C}^2\bo{X}_{[3,1]}$
by making use of the brane move \cite{FYY}.

\subsection{Seiberg--Witten periods}

Now our task is to calculate the Seiberg--Witten periods $a(z)$, $a_D(z)$ from
the torus periods through (\ref{eqn:period1})--(\ref{eqn:period3}).
The important subtlety arises in evaluating the integration constants.
First of all, since $\SW$ vanishes at $u=0$, $a(z)$ and $a_D(z)$ must vanish as
well at $z=0$ \cite{LMW}.
In the patch $|z|<1$, the SW periods are thus given by\footnote[4]{In the 
following we will ignore the irrelevant overall constants and put $R=1$ for 
simplicity. There is no difficulty in recovering them.} 
\begin{align}
a_D^{(0)}(z)&=
\int_0^z dz'\, \varpi_D^{(0)}(z'),\\
a^{(0)}(z)&=
\int_0^z dz'\, \varpi^{(0)}(z'). 
\end{align}
We note that these are succinctly expressed in terms of the generalized hypergeometric function 
\begin{equation}
{}_3 F_2(\alpha_1,\alpha_2, \alpha_3; \beta_1, \beta_2; z)=
\sum_{n=0}^{\infty}\frac{(\alpha_1)_n(\alpha_2)_n(\alpha_3)_n}
{(\beta_1)_n(\beta_2)_n}\frac{z^n}{n!}
\end{equation}
in such a way that
\begin{equation}
\left(\ba
a_D^{(0)}(z)\\a^{(0)}(z)
\ea\right)
=C_0\left(\ba
V_1^{(0)}(z)\\V_2^{(0)}(z)
\ea\right),
\label{eqn:period}
\end{equation}
where 
\begin{align}
V_1^{(0)}(z)&=\frac{1}{\alpha}\,z^{\alpha}\,{}_3F_2(\alpha, \alpha,\alpha\,;2\alpha, 1\!+\!\alpha\,;z),\\
V_2^{(0)}(z)&=\frac{1}{1-\alpha}\,z^{1-\alpha}\,{}_3F_2(1\!-\!\alpha,1\!-\!\alpha,1\!-\!\alpha\,;2(1\!-\!\alpha),2\!-\!\alpha;z).
\end{align}

Next the SW periods in the patch $|1-z|\leq 1$ are 
\begin{align}
a_D^{(1)}(z)&=
\int_1^z dz'\, \varpi_D^{(1)}(z')+c_D^{(1)},\\
a^{(1)}(z)&=
\int_1^z dz'\,\varpi^{(1)}(z')+c^{(1)},
\end{align}
where $c_D^{(1)}$ and $c^{(1)}$ are integration constants.
Notice that the analytic continuation allows us to write 
\begin{equation}
a_D^{(1)}(z)=\int_0^z dz'\, \varpi_D^{(1)}(z')
\end{equation}
which obeys $a_D^{(1)}(0)=a_D^{(0)}(0)=0$.
Therefore, 
\begin{align}
c_D^{(1)}&=a_D^{(1)}(1)\CR
&= -\pi\int_0^1 dx\, x^{-(1-\alpha)} {}_2F_1(\alpha,\alpha;1;1-x)\CR
&= -\pi\sum_{n=0}^{\infty}\frac{(\alpha)_n^2}{n!}
	\frac{\Gamma(\alpha)}{\Gamma(\alpha+n+1)}\CR
&= -\pi \frac{\Gamma(\alpha)}{\Gamma(\alpha+1)}\,
{}_2F_1(\alpha,\alpha;1+\alpha;1).
\end{align}
Here the well-known formula
\begin{equation}
{}_2F_1(\alpha,\beta;\gamma;1)
=\frac{\Gamma(\gamma)\Gamma(\gamma-\alpha-\beta)}{\Gamma(\gamma-\alpha)\Gamma(\gamma-\beta)},\quad
\mbox{Re}\,\gamma >0, \quad\mbox{Re}(\gamma-\alpha-\beta)>0
\end{equation}
helps us to find
\begin{equation}
c_D^{(1)}=-\pi\Gamma(\alpha)\Gamma(1-\alpha)=-\frac{\pi^2}{\sin\pi\alpha}.
\label{eqn:const}
\end{equation}
Since $c_D^{(1)}$ is also obtained from $c_D^{(1)}=a_D^{(0)}(1)$, 
(\ref{eqn:const}) yields the nontrivial identity for the generalized 
hypergeometric function ${}_3F_2$ at $z=1$ via (\ref{eqn:period}).
For $\alpha=\frac{1}{3}$, this identity was first found numerically in 
\cite{GL}, for which the above calculation affords analytic proof.

On the other hand, the similar integral for $a^{(1)}(1)$ is not helpful to 
determine $c^{(1)}$ analytically. Hence, evaluating $c^{(1)}=a^{(0)}(1)$
numerically we determine
\begin{equation}
c^{(1)}=\frac{\pi^2}{\sin\pi\alpha}\times
\begin{cases}
0.5000+0.9281i\quad \mbox{for}\; \wh{E}_8\\
0.5000+0.6103i\quad \mbox{for}\; \wh{E}_7\\
0.5000+0.4628i\quad \mbox{for}\; \wh{E}_6.
\end{cases}
\label{eqn:const2}
\end{equation}
These constants were first evaluated in \cite{MNW2,KZ}.\footnote[2]{We should 
note that our choice of the branch is $-z=e^{-i\pi}z$.
If the other branch $-z=e^{i\pi}z$ was chosen one would have $-0.5000$
instead of $0.5000$ in (\ref{eqn:const2}) in agreement with \cite{MNW2,KZ}.}

Let us now turn to the analysis of the SW periods around $z=\infty$.
In order to fix the integration constants, we adopt the idea in \cite{AGM}
and employ the Barnes' integral representation of the hypergeometric function 
\begin{equation}
{}_2F_1(\alpha,\beta;\gamma;z)=\frac{\Gamma(\gamma)}{\Gamma(\alpha)\Gamma(\beta)}
\int_{-i\infty}^{i\infty}\frac{dt}{2\pi i}
\frac{\Gamma(t+\alpha)\Gamma(t+\beta)\Gamma(-t)}{\Gamma(t+\gamma)}(-z)^t,
\label{eqn:barnes}
\end{equation}
where $|\arg(-z)|<\pi$. 
When the integration contour is closed on the right we have the power
series as presented in (\ref{eqn:bases}) which converges for $|z|<1$.
For our purpose, we first check how (\ref{eqn:barnes}) can be 
used to reproduce the SW periods in the patch $|z|<1$.
The naive idea is that expressing the torus periods in the form 
(\ref{eqn:barnes}), we first make the $z$-integral and then perform the 
contour integral with respect to $t$.
We are thus led to consider the integral 
\begin{equation}
I_{\alpha}(z)=\frac{\Gamma(2\alpha)}{\Gamma^2(\alpha)}
\int_{-i\infty}^{i\infty}\frac{dt}{2\pi i}\frac{\Gamma^2(t+\alpha)\Gamma(-t)}
{\Gamma(t+1-\alpha)(t+\alpha)}(-z)^t z^{\alpha}.
\end{equation}
Closing the contour on the right we immediately obtain 
\begin{equation}
I_{\alpha}(z)_{\text{right}}=\frac{1}{\alpha}z^{\alpha}\,
{}_3F_2(\alpha,\alpha,\alpha;2\alpha,1+\alpha;z),
\end{equation}
and hence (\ref{eqn:period}) is reproduced.

If we close the contour on the left, we obtain the 
expression which is valid for $|z|>1$. Thus
\begin{equation}
\left(\ba
a_D^{(\infty)}(z)\\a^{(\infty)}(z)
\ea\right)
=C_0\left(\ba
I_{\alpha}(z)_{\text{left}}\\
I_{1-\alpha}(z)_{\text{left}}
\ea\right).
\end{equation}
Upon doing the contour integral on the left, 
one has to pick up the triple pole at $t=-\alpha$
in addition to the standard poles in the Barnes' integral.
This in fact gives rise to the integration constant.
After some algebra we arrive at 
\begin{equation}
\left(\ba
a_D^{(\infty)}(z)\\a^{(\infty)}(z)
\ea\right)
=C_{\infty}\left(\ba
V_1^{(\infty)}(z)\\ V_2^{(\infty)}(z)\ea\right)
+C_0\left(\ba b_1\\b_2\ea\right),
\label{eqn:infty}
\end{equation}
where
\begin{align}
V_1^{(\infty)}(z)&=\log(-z)-g(z),\\
V_2^{(\infty)}(z)&=\frac{1}{2}\log^2(-z)-\log(-z)g(z)+h(z),
\end{align}
with 
\begin{align}
g(z)&=\sum_{n=1}^{\infty}\frac{(\alpha)_n(1-\alpha)_n}{(n!)^2 n}\frac{1}{z^n},
\\
h(z)&=\sum_{n=1}^{\infty}\frac{(\alpha)_n(1-\alpha)_n}{(n!)^2 n}
\left[\sum_{k=0}^{n-1}\left(\frac{1}{\alpha+k}+\frac{1}{1-\alpha+k}
-\frac{2}{1+k}\right)-\frac{1}{n}\right]\frac{1}{z^n}.
\end{align}
In (\ref{eqn:infty}) the integration constants read 
\begin{equation}
b_1=\frac{1}{2}e^{i\pi\alpha}\xi_1\left(\eta_1^2+\frac{\pi^2}{3}\right),
\quad
b_2=\frac{1}{2}e^{i\pi(1-\alpha)}\xi_2\left(\eta_2^2+\frac{\pi^2}{3}\right).
\end{equation}
Further manipulations yield
\begin{equation}
a_D^{(\infty)}(z)=-4\pi^2\sin\pi\alpha\,\frac{H_2(z)}{(2\pi i)^2},\quad
a^{(\infty)}(z)=\frac{\pi^2}{\sin\pi\alpha}\frac{H_1(z)}{2\pi i},
\label{eqn:rel}
\end{equation}
where
\begin{align}
H_1(z)&=-\log(-vz)+g(z),\\
H_2(z)&=-\frac{1}{2}\log^2(vz)+\log(vz)g(z)-h(z)-\frac{\pi^2}{2}
\left(\frac{1}{\sin^2\pi\alpha}+\frac{1}{3}\right).
\end{align}

Now that we have fixed all the integration constants, it is seen that
the monodromy matrices with integral entries are obtained by setting the 
constant period 
\begin{equation}
s=\frac{\pi^2}{\sin\pi\alpha}.
\end{equation}
As a result, the monodromy matrices acting on the period vector 
${}^t\Pi={}^t(s,a(z),a_D(z))$ for $\wh{E}_{N=8,7,6}$ theories turn out to be
\begin{align}
M_0=\begin{pmatrix}
1&0&0\\
0&1&1\\
0&N-9&N-8
\end{pmatrix},
\quad
M_1=\begin{pmatrix}
1&0&0\\
-1&1&-1\\
0&0&1
\end{pmatrix},
\quad
M_{\infty}=\begin{pmatrix}
1&0&0\\
-1&1&0\\
9-N&N-9&1
\end{pmatrix}\CR 
\end{align}
which obey $M_{\infty}=M_0M_1$, $M_0^6={1}$ ($\wh{E}_8$), 
$M_0^4={1}$ ($\wh{E}_7$) and $M_0^3={1}$ ($\wh{E}_6$).

\begin{figure}
\hspace{1.5cm}
\epsfxsize=13cm
\epsfbox{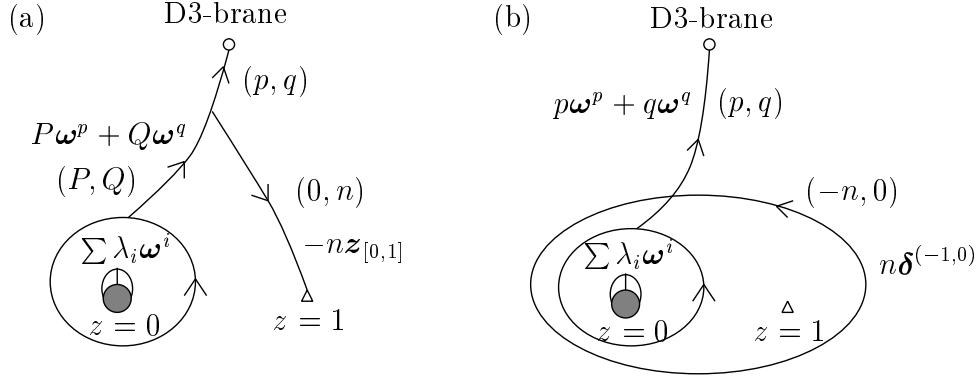}
\caption{String junctions on the $\wh{E}_N$ 7-brane configuration}
\label{fig:junction}
\end{figure}

Finally the BPS central charge is expressed in terms of the period 
integrals
\begin{equation}
Z=pa(z)-qa_D(z)+ns,
\label{eqn:cc}
\end{equation}
where $(p,q)\in \bo{Z}^2$ are electric and magnetic charges and 
$n\in \bo{Z}$ is global conserved charge of the BPS states.
In view of the D3-probe picture, (\ref{eqn:cc}) is the central charge 
corresponding to the string junction,
\begin{equation}
\bo{J}=\sum_{i=1}^{N}\lambda_i \bo{\omega}^i+P\,\bo{\omega}^p+
Q\,\bo{\omega}^q-n\,\bo{z}_{[0,1]},
\label{eqn:junction}
\end{equation}
where $P=p$, $Q=q+n$ and our notation is as follows;
$\{\lambda_i\}$ is the Dynkin label of $E_N$ weights and the 
$\bo{\omega}^i$ are the junctions of zero asymptotic charges 
representing the fundamental weights of the Lie algebra $E_N$,
$\bo{\omega}^p$, $\bo{\omega}^q$ are the $E_N$ singlet junctions with 
asymptotic charges $(1,0)$, $(0,1)$, respectively, 
and $\bo{z}_{[0,1]}$ stands for the outgoing (0,1)-string emanating from 
the 7-brane $\bo{X}_{[0,1]}$ at $z=1$.
See Fig.~\ref{fig:junction}a for the $\wh{E}_N$ brane-junction configurations 
probed by a single D3-brane.
To justify the above correspondence, for instance, take a junction with 
charges $(p,q\,;n)=(0,-1 ;1)$. 
This is a single BPS string stretched between the D3-brane and the 7-brane
$\bo{X}_{[0,1]}$.
Clearly its mass vanishes when the D3-brane is located at the point $z=1$.
This is indeed verified from (\ref{eqn:cc}) 
since $Z(z)=a_D(z)+s$ for $(0,-1;1)$ charges and $Z(1)=a_D(1)+s=0$
at $z=1$ by virtue of (\ref{eqn:const}).

We note that there are no terms in (\ref{eqn:cc}) which reflect the presence
of the $E_N$ non-singlet junctions in (\ref{eqn:junction}) explicitly.
To have such terms one has to turn on mass deformation parameters in the
$\wh{E}_N$ curves so that the SW differentials have poles with 
non-vanishing residues other than the one at $x=0$ ($x=iu/R$ for $\wh{E}_6$). 
This will produce additional terms in (\ref{eqn:cc}) which may depend 
directly on the $E_N$ representations.

Notice that the string junction (\ref{eqn:junction}) is equivalently 
expressed as 
\begin{equation}
\bo{J}=\sum_{i=1}^{N}\lambda_i \bo{\omega}^i+p\,\bo{\omega}^p+
q\,\bo{\omega}^q+n\,\bo{\delta}^{(-1,0)},
\label{eqn:junction2}
\end{equation}
where $\bo{\delta}^{(-1,0)}$ is a loop junction which represents the 
imaginary root of the affine algebra $\wh{E}_N$, see Fig.~\ref{fig:junction}b.
In the junction realization of $\wh{E}_N$,
the level $k$ of representations is given by 
$k(\bo{J})=-(\bo{J},\bo{\delta}^{(-1,0)})=-q$, 
while the grade $\bar{n}(\lambda)$ for a weight vector 
$\{\lambda_i\}$ is equal to $n$ in $\bo{J}$ up to a constant 
shift \cite{DHIZ}.

On the D3-brane probing the region $|z|<1$, in the limit $R\rightarrow 0$, 
the theory reduces to the 4D superconformal theory with global exceptional 
symmetries \cite{MN,NTY}.
Recovering the $R$-dependence we have the central charge in the form 
\begin{equation}
Z=pa(u)-qa_D(u)+\frac{n}{R}s,
\end{equation}
where $a(u)$, $a_D(u)\simeq \mbox{const}\cdot u^{\alpha}$ and the
last term represents the KK modes associated with the $S^1$ compactification.
We thus observe that the grading of KK modes is regarded as the the grade
in the affine symmetry $\wh{E}_N$.
As will be seen in section~\ref{sec:map} these KK modes are identified 
with the D0-branes in M-theory.
The decoupling of the KK modes as $R\rightarrow 0$ is ensured 
since the 7-brane $\bo{X}_{[0,1]}$ moves away to infinity on the $u$-plane, 
and hence the loop junction $\bo{\delta}^{(-1,0)}$ decouples, leaving
the finite symmetry $E_N$ for 4D theories.

%%%%%%%%%%%%%%%%
\section{Mirror map}
%%%%%%%%%%%%%%%%
\renewcommand{\theequation}{3.\arabic{equation}}\setcounter{equation}{0}

5D $\cN$=1 supersymmetric $SU(2)$ gauge theories with 
exceptional global symmetries \cite{S} are realized by compactifying M-theory
on a \cy threefold with a vanishing del Pezzo 
four-cycle \cite{MS,DKV}.
In \cite{LMW} the local mirror geometry of the singularity associated with a
shrinking del Pezzo four-cycle of the type $E_{N=8,7,6}$ is modeled by the
Landau-Ginzburg potential
\begin{align}
W_{E_8}&=\frac{1}{x_0^6}+x_1^2+x_2^3+x_3^6+x_4^6-\psi x_0x_1x_2x_3x_4,\CR
W_{E_7}&=\frac{1}{x_0^4}+x_1^2+x_2^4+x_3^4+x_4^4-\psi x_0x_1x_2x_3x_4,
\label{eqn:delpezzo}\\
W_{E_6}&=\frac{1}{x_0^3}+x_1^3+x_2^3+x_3^3+x_4^3-\psi x_0x_1x_2x_3x_4.\nonumber
\end{align}
The equations $W_{E_N}=0$ describe non-compact \cy manifolds and $\psi$
is a complex moduli parameter.
The point $\psi=0$ is the $E_N$ symmetric point (the Landau-Ginzburg
point) at which the del Pezzo four-cycle collapses.
The large complex structure limit is taken by letting $\psi=\infty$.

It turns out that the natural complex modulus is $\psi^{\ell}$ with 
$\ell=6$ ($E_8$), 4 ($E_7$) and 3 ($E_6$) and, as will be seen momentarily, 
$\psi^{\ell}$ and $z$ in the previous section is related through
\begin{equation}
z=\frac{\psi^6}{432}\;(E_8),\quad
\frac{\psi^4}{64}\;(E_7),\quad
\frac{\psi^3}{27}\;(E_6).
\label{eqn:parameter}
\end{equation}
The period integral $\wt{\Pi}$ of the holomorphic three-form $\Omega$
associated with (\ref{eqn:delpezzo}) are defined over a basis of three-cycles
on the mirror \cy\!\!\!.
They obey the \PF equations \cite{LMW}
\begin{equation}
{\cal L}_{\text{ell}}^{E_N}\cdot \vartheta \,\wt{\Pi}=0,
\label{eqn:PF2}
\end{equation}
where $\vartheta=c\frac{\partial}{\partial c}$ with $c=\psi^{-\ell}$ and
${\cal L}_{\text{ell}}^{E_N}$ are the \PF operators
corresponding to the elliptic singularities of type $\wh{E}_N$
\begin{align}
{\cal L}_{\text{ell}}^{E_8}&=
\vartheta^2-12 c (6\vartheta+5)(6\vartheta+1),\CR
{\cal L}_{\text{ell}}^{E_7}&=
\vartheta^2-4c(4\vartheta+3)(4\vartheta+1),\\
{\cal L}_{\text{ell}}^{E_6}&=
\vartheta^2-3c(3\vartheta+2)(3\vartheta+1).\nonumber
\end{align}
Making a change of variables with (\ref{eqn:parameter}), one can
check that (\ref{eqn:PF2}) is equivalent to (\ref{eqn:PF1}).
In fact, it is shown in \cite{LMW} that the \cy periods 
$\wt{\Pi}(c)$ over three-cycles reduce to the SW periods $\Pi$ after
performing the integrals over appropriate two-cycles. 
Thus we have $\wt{\Pi}=\Pi=(s,a(z),a_D(z))$.

Under mirror symmetry three-cycles on the IIB side is mapped to the 
zero, two, four-cycles on the IIA side where the four-cycle is the del 
Pezzo surface.
Following the standard machinery \cite{CDGP} we find that the complex K\"ahler
modulus $t$ is given by
\begin{equation}
t(z)=\frac{H_1(z)}{2\pi i}
\end{equation}
and its dual $t_d$ becomes 
\begin{equation}
t_d(z)=\frac{\partial {\cal F}}{\partial t}=-\frac{H_2(z)}{(2\pi i)^2}.
\end{equation}
Thus the SW periods and the flat coordinate system are related via
\begin{equation}
\left(\ba
s\\a(z)\\a_D(z)
\ea\right)
=\frac{\pi^2}{\sin\pi\alpha}\left(\ba
1&0&0\\
0&1&0\\
0&0&4\sin^2\pi\alpha
\ea\right)
\left(\ba
1\\t(z)\\t_d(z)
\ea\right).
\end{equation}
The central charge (\ref{eqn:cc}) is then rewritten as
\begin{equation}
Z=\frac{\pi^2}{\sin\pi\alpha}
\left(-4\sin^2\pi\alpha\,q\,t_d(z)+p\,t(z)+n\right).
\label{eqn:cc2}
\end{equation}

In the large radius region $|z|>1$, it is shown from (\ref{eqn:rel})
that
\begin{equation}
t_d=\frac{t^2}{2}-\frac{t}{2}
-\frac{1}{24}\left(\frac{3}{\sin^2\pi\alpha}-2\right)+O(e^{2\pi it}),
\label{eqn:lr}
\end{equation}
{}from which we see that $t_d$ is the central charge of the D4-brane.
Integrating this over $t$ we obtain the prepotential
\begin{equation}
{\cal F}=\frac{t^3}{6}-\frac{t^2}{4}
-\frac{1}{24}\left(\frac{3}{\sin^2\pi\alpha}-2\right)t+\mbox{const.}
-\frac{1}{(2\pi i)^3}\sum_{k=1}^{\infty}n_k Li_3(e^{2\pi ikt}),
\end{equation}
where $Li_3(x)$ is the trilogarithm and the instanton coefficients $n_k$
are presented in Table \ref{tbl2}.
We note that the $n_k$'s multiplied by $4\sin^2\pi\alpha$ for
$\wh{E}_N$ coincide with the Gromov--Witten invariants obtained in 
\cite{KMV}.
\begin{table}
\begin{center}
\begin{tabular}{|c|r|r|r|} 
\hline
	&$\wh{E}_8$   		&$\wh{E}_7$  		&$\wh{E}_6$\\	
\hline
$n_1$	&$252$			&$28$			&$9$\\
$n_2$	&$-9252$		&$-136$			&$-18$\\
$n_3$	&$848628$		&$1620$			&$81$\\
$n_4$	&$-114265008$		&$-29216$		&$-576$\\
$n_5$	&$18958064400$		&$651920$		&$5085$\\
$n_6$	&$-3589587111852$	&$-16627608$		&$-51192$\\
$n_7$	&$744530011302420$	&$465215604$		&$565362$\\
$n_8$	&$-165076694998001856$	&$-13927814272$		&$-6684480$\\
$n_9$	&$38512679141944848024$&$439084931544$		&$83246697$\\
$n_{10}$&$-9353163584375938364400$&$-14417814260960$	&$-1080036450$\\
\hline
\end{tabular}
\end{center}
\caption{Instanton coefficients}
\label{tbl2}
\end{table}

%%%%%%%%%%%%%%%%
\section{D-branes on a surface}
%%%%%%%%%%%%%%%%
\renewcommand{\theequation}{4.\arabic{equation}}\setcounter{equation}{0}

The aim of this section is to obtain 
several basic invariants of BPS D-branes bounded on a surface, which are
expressed as  coherent sheaves on the surface,  
such as the RR charge, the central charge,
and some intersection pairings with the aid of 
some algebraic geometry techniques.
They play a crucial role in the duality map between D-branes on 
del Pezzo surfaces and the string junctions with $\EN$ symmetry,
which we will discuss in the next section.
An important note is that though we identify a BPS D-brane on a surface
with a \coh sheaf, we do not enter into the conditions, such as the stability,
to be satisfied by the sheaf to represent a true BPS brane, 
with few exceptions; in effect,
what we really need is not a sheaf itself but
its invariants such as the Chern character in this article.

\subsection{D-branes on a Calabi--Yau threefold}
We represent a BPS D-brane
on a \cy threefold $X$ by a \coh $\cO_X$-module $\cG$.
The RR charge %(or RR charge in short) 
of $\cG$ is given by the Mukai vector \cite{Muk,HM}
\begin{equation}
v_X\left(\cG\right)=
\ch\left(\cG\right)\sqrt{\Todd\left(T_X\right)}
\in H_{2\bullet}(X;\bo{Q})
:=\bigoplus_{i=0}^{3}H_{2i}\left(X;\bo{Q}\right),
\end{equation}
where 
$\ch(\cG)=\sum_{i=0}^3\ch_i(\cG)$
%\ch_0(\cG)+\ch_1(\cG)+\ch_2(\cG)+\ch_3(\cG)$
is the Chern character with $\ch_i(\cG)\in H_{6-2i}(X;\bo{Q})$,
which can be computed as follows:
there always exits a resolution of $\cG$ by locally free sheaves 
 $(V_{\bullet})$, that is, sheaves of sections of 
holomorphic vector bundles:
$0\to V_3\to V_2\to V_1 \to V_0 \to \cG\to 0$ (exact), 
thus we can set $\ch(\cG)\colon=\sum_{i=0}^3\,(-1)^i\,\ch(V_i)$,
which does not depend on the choice of the resolution,
and $\sqrt{\Todd(T_X)}=[X]+c_2(X)/24$, the effect of which on the
RR charges has been called a geometric version 
of the Witten effect in \cite{HM}.

The intersection form on D-branes on $X$,
which is of great importance in investigating
topological aspects of D-branes \cite{Gepner,Gepner2}, 
is given by
\begin{equation}
I_X\left(\cG_1,\cG_2\right)
=\big\langle v_X(\cG_1)^{\vee}\cdot v_X(\cG_2)\big\rangle_X=
\big\langle \ch\left(\cG_1\right)^{\vee}\cdot
\ch\left(\cG_2\right)\cdot\Todd
\left(T_X \right)\big\rangle_X,
\label{intX}
\end{equation}
where $\langle \cdots \rangle_X$ evaluates the degree of 
$H_0\left(X;\bo{Q}\right)\cong \bo{Q}$ 
component, 
%which can also be denoted as $\int_X \cdots$,
and $v\rightarrow v^{\vee}$ 
flips the sign of $H_{0}(X)\oplus H_4(X)$ part.
In particular, if $\cG$ itself is locally free, then 
$\ch(\cG)^{\vee}=\ch(\cG^{\vee})$, 
where 
$\cG^{\vee}={\cal H}om_{X}(\cG,\cO_X)$
is the dual sheaf.

It is easy to check 
$I_X(\cG_2,\cG_1)=-I_X(\cG_1,\cG_2)$.
On the other hand, the Hirzebruch--Riemann--Roch formula \cite{Hir,HL} 
tells us   
\begin{equation}
I_X\left(\cG_1,\cG_2\right)
=\sum_{i=0}^3\,(-1)^i\,{\text{dim}}\,{\text{Ext}}^i_X(\cG_1,\cG_2),
\label{Ext}
\end{equation}
according to which the skew-symmetric property 
of the intersection form $I_X$ may be attributed  
to the Serre duality:  
${\text{Ext}}_X^i(\cG_2,\cG_1)
\cong{\text{Ext}}_X^{3-i}(\cG_1,\cG_2)^{\vee}$ \cite{ShH}.
Incidentally, the H.R.R.~formula
(\ref{Ext}) also assures that $I_X$ takes values in $\bo{Z}$.

Let $J_X\in H_4(X;\bo{R})$ be a \K form on $X$, which is 
identified with an $\bo{R}$-extended ample divisor here. 
The classical expression of the central charge 
of the D-brane $\cG$ is then given by \cite{Gepner,Gepner2}
\begin{equation}
Z_{J_X}^{\text{cl}}(\cG)
=-\big\langle e^{-J_X}\cdot v_X(\cG)\big\rangle_X
=-\sum_{k=0}^3 \frac{(-1)^{k}}{k!}
\big\langle J^k_X \cdot v_{X,k}(\cG) \big\rangle_X,
\label{classicalX}
\end{equation}
where $v_{X,k}$ is the $H_{2k}(X)$ component of 
$v_X\in H_{2\bullet}(X)$.
On the other hand, the quantum central charge
$Z_{J_X}(\cG)$ differs from its classical counterpart (\ref{classicalX})
by the terms of order  
$O(e^{-2\pi \bra J_X\cdot \beta  \ket_X})$
where two-cycle $\beta$ is in the Mori cone of $X$, 
which is dual to the \K cone; 
the exact \K moduli dependence of $Z_{J_X}$ can be  
determined in principle by the \PF equations for the periods 
of the mirror \cy threefold  $X^{\vee}$ \cite{GK,GL,GL2}.

%% %% %% %% %% %% %% %% %% %% %% %% %% %% %% %% 
\subsection{D-branes localized on a surface}
%% %% %% %% %% %% %% %% %% %% %% %% %% %% %% %%  
Let $f\colon S\,{\hookrightarrow}\,X$ be an embedding of a projective
surface $S$ in a \cy threefold $X$.
If $c_1(S)\in H_2(S)$ is nef, which means 
that its intersection with any effective curve $C$ on $S$
are non-negative: $\cfS\cdot C\geq 0$,  
there is a smooth elliptic \cy threefold over $S$ with $S$ 
its cross section, which we can take as a model of embedding \cite{CKYZ}.
Other examples of embedding can be found in \cite{LMW,HM,Toshiya,HST}.

Now let us take the limit of infinite elliptic fiber,
so that the D-branes the central charge 
of which remains finite are those which are confined to the surface $S$,
where we should note that some D-branes on $S$,
a D0-brane for example, can move along elliptic fibers so as to leave $S$
even if $S$ itself is rigid in $X$.
The properties of the D-branes localized on $S$ 
then depend not on the details of the global model $X$,
but only on the intrinsic geometry of $S$ and its normal bundle
$N_S=N_{S|X}$, which is isomorphic to the canonical line bundle $K_S$.
In particular, this means that we can compute the central charges 
of BPS D-branes using local mirror symmetry principle on $S$ \cite{CKYZ}.

A D-brane sticking to $S$ can be described by a \coh 
$\cO_S$-module $\cE$.  
The Euler number of it defined by
$\chi(\cE)=\sum_{i=0}^2\,(-1)^i\,h^i(S,\cE)$, where
$h^i(S,\cE)={\text{dim}}\,H^i(S,\cE)$,
is an important invariant, which can be obtained  as follows:
first we need the Todd class of $S$
\begin{align}
\Todd(T_S)&=[S]+\frac12\,c_1(S)+\chi(\cO_S)\,[{\text{pt}}],\\
\chi(\cO_S)&=\frac{1}{12}\big\bra \csS+\cfS^2 \big\ket_S,
\end{align}
second, by the H.R.R.~formula, we have
\begin{equation}
\chi(\cE)=\big\bra \ch(\cE)\,\Todd(T_S)\big\ket_S
=r\, \chi(\cO_S)
+\big\bra \ch_2(\cE)+\frac12\,c_1(S)\cdot c_1(\cE) \big\ket_S.
\label{EulercE}
\end{equation}

%%%%%
There is a canonical push-forward homomorphism 
$f_*\colon H_{2\bullet}(S;\bo{Q})\,{\rightarrow}\,H_{2\bullet}(X;\bo{Q})$,
which maps a cycle on $S$ to that on $X$.
Similarly, we can define the \coh sheaf $f_!\,\cE$
on $X$ by extending $\cE$ by zero to $X\backslash S$.%
\footnote[3]{The symbol $f_!\,\cE$  is originally defined to be 
$\sum_i(-\!1)^i R^i\!f_*\,\cE$, an element of  
the K group of \coh $\cO_X$-modules \cite{BS,Hir};
it reduces to the single direct image sheaf $f_*\,\cE$ on $X$,
because all the higher direct images $R^i\!f_*\,\cE$ vanish 
for embedding $f$ \cite[p.~102]{BS}.}
The celebrated Grothendieck--Riemann--Roch formula \cite{Hir,BS}
for embedding $f\colon S\,{\hookrightarrow}\,X$ 
relates the Chern characters of $\cE$ and $f_!\, \cE$ as follows:
\begin{equation}
\ch(f_!\,\cE)=f_*\left(\ch(\cE)\,\frac{1}{\Todd(N_S)}\right).
\label{GRR}
\end{equation}
%
%where   
%$f_*\colon H_{2\bullet}(S;\bo{Q})\,{\rightarrow}\,H_{2\bullet}(X;\bo{Q})$
%is the homomorphism,
%while by $f_!\,\cE$ we mean the same $\cE$ 
%now regarded as a sheaf on $X$, which has its support on $S$.
%
Multiplying the both hand side of (\ref{GRR}) by the square root of
$\Todd(T_X)$, we have 
\begin{equation}
\ch(f_!\,\cE)\sqrt{\Todd(T_X)}=
f_*\left(\ch(\cE)\sqrt{\frac{\Todd(T_S)}{\Todd(N_S)}}\right),
\label{bcS}
\end{equation}
where we have used the projection formula 
\cite[p.~273]{Da}, \cite[p.~426]{Ha}:
\begin{equation} 
f_*\left(\alpha\cdot\!f^*\beta\right)=f_*\alpha\cdot\!\beta,
\qquad \alpha\in H_{2\bullet}(S;\bo{Q}),\quad
 \beta\in H_{2\bullet}(X;\bo{Q}), 
\label{projection}
\end{equation}
and  $f^*\Todd(T_X)=\Todd(T_S)\cdot\Todd(N_S)$, 
which follows from the short exact sequence of bundles on $S$:
$
0\rightarrow T_S\rightarrow f^* T_X\rightarrow N_S\rightarrow 0,
$
combined with the multiplicative property of the Todd class.
As the left hand side of (\ref{bcS}) is
the D-brane charge of $\cE$ regarded as a brane on $X$,
we arrive at the intrinsic description of 
the RR charge on $S$:
%\footnote{In (\ref{RRonS}), we have supressed the
%contribution of the NS B-field $e^{-\frac{B}{2\pi}}$.
%On the other hand, $S$ admits a spin structure if and only if 
%$c_1(S)
%
\begin{equation}
v_S(\cE)=\ch(\cE)\sqrt{\frac{\Todd(T_S)}{\Todd(K_S)}}
=\ch(\cE)\,
e^{\frac12 \cfS}\,
\sqrt{\frac{\widehat{A}(T_S)}{\widehat{A}(K_S)}}
\in H_{2\bullet}(S;\bo{Q}),
\label{RRonS}
\end{equation}
which is a complex-analytic (or algebraic)
derivation of the RR charge which has originally been 
obtained in $C^{(\infty)}$ category \cite{Moore,CY}.
The gravitational correction factor for $S$
admits the following expansion:
\begin{equation}
\sqrt{\frac{\Todd(T_S)}{\Todd(K_S)}}=[S]+\frac12 \cfS
+\frac{1}{24}\big( \csS+3\cfS^2 \big)
\in H_{2\bullet}(S;\bo{Q}).
\label{gravi}
\end{equation}
%
%where we have used $\cfS^2=12-\chi(S)$. 
%\vspace{0.5cm}

As a simple exercise let us compute the RR charge of a sheaf on $S$.
To this end, let $\iota\colon C\hookrightarrow S$ be an embedding of 
a smooth genus $g$ curve in $S$ with normal bundle 
$N_C=N_{C|S}$.
Then from a line bundle $L_C$ on $C$,
we obtain a torsion sheaf
%-------------------------------
%\footnote{A sheaf is called torsion if its stalk at the generic point
%is zero.}%
%-------------------------------
$\iota_! L_C$ on $S$.
$\ch(\iota_!\,L_C)$ can be computed again from the G.R.R.~formula:
\begin{equation}
\ch(\iota_!\,L_C)=\iota_*
\left(\ch(L_C)\, \frac{1}{\Todd(N_C)}\right)
=\iota_*[C]+\big(\deg(L_C)-\frac12\deg(N_C)\big)[{\text{pt}}],
%\in H_2(S;\bo{Q})\oplus H_0(S;\bo{Q}),
\end{equation}
where $\deg(L):=\bra c_1(L)\ket_C$ for a line bundle $L$ on $C$.
The RR charge of the torsion
$\cO_S$-module $\iota_!\,L_C$ can then be computed as 
\begin{equation}
v_S(\iota_!\,L_C)=\iota_*[C]
%+\left(\deg(L_C)-g+1\right)[{\text{pt}}]
+\chi(L_C)[{\text{pt}}]
\in H_2(S)\oplus H_0(S),
\label{curve}
\end{equation}
where we have used the classical adjunction and 
the self-intersection formulae on $S$:
\begin{align}
2g(C)-2&=\big\bra [C]\cdot [C]-[C]\cdot \cfS\big\ket_S,
\label{adjunction}\\
\deg(N_C)&=\big\bra [C]\cdot [C]\big\ket_S,
\end{align}
as well as the classical Riemann--Roch formula on $C$:
\begin{equation}
\chi(L_C)=h^0(C,L_C)-h^1(C,L_C)=\deg(L_C)+1-g.
\end{equation}
%
%According to the classical Riemann--Roch formula,
%the D0-brane charge $\deg(L_C)+1-g$ in (\ref{curve})
%is just the Euler number of the line bundle
%$\chi(L_C)=h^0(C,L_C)-h^1(C,L_C)$.
%The computation given here will be used to map the RR charge of 
%a D2-brane wrapped on $C$ to  a string junction 
%in the next section (\ref{Dtwo}).

%--------------------------------------------------------------

Let us next turn to intersection pairings on D-branes.
It seems that the most appropriate intersection form
on D-branes on $S$ {\em could} depend on one's purpose. 
Below, we will describe three candidates, each of which we think 
has its own reason to be chosen as an intersection form. 

%\\\\\\\\\\\\\\\\\\\\\\\\\\\\\\\\\\\\\\\\\\\\\\\\\\\\\\\\
The first uses the Mukai vector $v_S$ (\ref{RRonS}) and defines 
a symmetric form:
\begin{align}
I_S(\cE_1,\cE_2)
&=-\big\bra 
v_S(\cE_1)^{\vee}\cdot v_S(\cE_2)
\big\ket_S  \nonumber \\
&= \frac{1}{12}\,r_1r_2\,\chi(S)
+\big\bra r_1\, \ch_2(\cE_2)+r_2\, \ch_2(\cE_1)
-c_1(\cE_1)\cdot c_1(\cE_2)\big\ket_S,
\label{intS1}
\end{align}
where $\ch(\cE)=r[S]+c_1(\cE)+\ch_2(\cE)$, 
$\chi(S)=\bra \csS\ket_S$ the Euler number of $S$,
and $v^{\vee}=-v_0+v_1-v_2$ \cite{Muk,HL},
with $v_i$ being the $H_{2i}(S)$ component of $v$.
It should also be noted that $I_S$ is {\em not} $\bo{Z}$-valued
in general.
%\\\\\\\\\\\\\\\\\\\\\\\\\\\\\\\\\\\\\\\\\\\\\\\\\\\\\\\\\\\

The second is the skew-symmetric form $I_X\scirc f_!$
induced from the one on the ambient \cy threefold $X$ (\ref{intX}).
As shown below, however, this form has an  description intrinsic to $S$
independent of the details of embedding:
\begin{align}
I_X(f_!\,\cE_1,f_!\,\cE_2)
&=\big\bra f_*(v_S(\cE_1))^{\vee} \cdot f_*(v_S(\cE_2)) \big\ket_X 
\nonumber \\
&=\big\bra r_1\,c_1(\cE_2)\cdot \cfS
-r_2\,c_1(\cE_1)\cdot \cfS\big\ket_S,
\label{intS2}
\end{align}
where we have used 
the self-intersection formula:
$f^*f_*[S]=c_1(N_S)$ \cite[p.~431]{Ha},
as well as the projection formula (\ref{projection}) 
to show that for $[S]\in H_4(S)$, $[C]\in H_2(S;\bo{Q})$
\begin{equation}
f_*[S]\cdot f_*[C]=f_*\big([C]\cdot f^*f_*[S]\big)
=-f_*\big([C]\cdot \cfS\big).
\end{equation}
%\\\\\\\\\\\\\\\\\\\\\\\\\\\\\\\\\\\\\\\
To be explicit, consider $S=\bP^2$. $H_2(\bP^2)$ is 
isomorphic to $\bo{Z}$, the ample generator of which we denote by $l$.
Then $c_1(\bP^2)=3l$, and $\bra l\cdot l\ket_{\bP^2}=1$. 
Following Diaconescu and Gomis \cite{DG}, we express the Chern character 
of a \coh sheaf $\cE$ on $\bP^2$ as
\begin{equation}
r(\cE)=-n_2,\quad  c_1(\cE)=n_1 l,\quad  \ch_2(\cE)=-n_0[{\text{pt}}],
\end{equation} 
where $n_1,n_2\in \bZ$, and  $n_0\in \frac12 n_1+\bZ$.
Our second intersection form can be written in these variables 
as follows:
\begin{equation}
%I_X(f_!\,\cE,f_!\,\cE^{'})=-3(n_1^{'}n_2-n_1n_2^{'}),
I_X(f_!\,\cE,f_!\,\cE')=-3(n_1'n_2-n_1n_2'),
\label{DiaGom}
\end{equation}
which is precisely the intersection form in \cite{DG}
up to sign.
An interesting  remark is that
the intersection form (\ref{DiaGom}) introduced in \cite{DG} is  
based on that of one-cycles on an $\widehat{E}_6$ torus {\em contained} in
$\bP^2$, while our $I_X\scirc f_!$ is induced from 
that on a \cy threefold $X$ which {\em contains} $\bP^2$.
%\\\\\\\\\\\\\\\\\\\\\\\\\\\\\\\\\\\\\\\

%\\\\\\\\\\\\\\\\\\\\\\\\\\\\\\\\\\\\\\\\\\\\\\\\\\\\\\\\\\\\\
The third, which  has been used in \cite{HI}
to identify the RR charge lattice $H_{2\bullet}(S)$ 
with $S$ a \dP surface and the string junction charge lattices,
generalizing the result in \cite{DHIK},
would  be the most natural one also from 
the mathematical point of view \cite{HL,Muk}: 
\begin{align}
\chi_S(\cE_1,\cE_2)
&=\sum_{i=0}^{2}(-1)^i\dim{\text{Ext}}_S^i(\cE_1,\cE_2)
\nonumber \\
&=-\big\bra 
\ch(\cE_1)^{\vee}\cdot \ch(\cE_2)\cdot \Todd(T_S)
\big\ket_S \nonumber \\
&=r_1r_2\,\chi(\cO_S)
+\big\bra r_1\,\ch_2(\cE_2)+r_2\,\ch_2(\cE_1)
-c_1(\cE_1)\cdot c_1(\cE_2)\big\ket_S
\nonumber \\
&+\frac12\,\big\bra r_1\, c_1(\cE_2)\cdot \cfS
-r_2\, c_1(\cE_1)\cdot \cfS\big\ket_S.      
\label{most-natural-one}
\end{align}
%
%\\\\\\\\\\\\\\\\\\\\\\\\\\\\\\\\\\\\\\\\\\\\\\\\\\\\\\\\\\\\\
The skew-symmetric part of the third form 
$\chi_S$ coincides with $I_X\scirc f_!$:
\begin{equation}
\chi_S(\cE_1,\cE_2)-\chi_S(\cE_2,\cE_1)
=I_X(f_!\,\cE_1,f_!\,\cE_2),
\label{anti-sym}
\end{equation}
while the relation between the symmetric part 
of $\chi_S$ and $I_S$ becomes 
\begin{equation}
\frac{1}{2}\big(\chi_S(\cE_1,\cE_2)+\chi_S(\cE_2,\cE_1)\big)
=\frac{1}{12}\, r_1r_2 \left\bra c_1(S)^2 \right\ket_S
+I_S(\cE_1,\cE_2).
\end{equation}
In view of the Serre duality:
${\text{Ext}}_S^i(\cE_1,\cE_2)\cong
{\text{Ext}}_S^{2-i}(\cE_2,\cE_1\otimes K_S)^{\vee}$,
the skew-symmetric part of $\chi_S$ (\ref{anti-sym}) comes from 
the non-triviality of the canonical line bundle $K_S$.
%

%According to the Bogomolov ienequality \cite{HL},
According to the Bogomolov inequality,
the discriminant of a sheaf $\cE$, defined by
$$
\Delta(\cE):=\big\bra -2r\,\ch_2(\cE)+c_1(\cE)^2\big\ket_S,
$$
must be non-negative if $\cE$ is torsion-free%
%----------------------------
\footnote[4]{
Roughly speaking, 
a torsion-free sheaf on surface $S$ is a sheaf of sections of 
a vector bundle  with at worst point-like singularities.}%
%---------------------------- 
and semi-stable \cite{HL},
which puts the following constraint 
on the self-intersection number of a torsion-free $\cO_S$-module
$\cE$ corresponding to a true BPS D-brane:
\begin{equation}
\chi_S(\cE,\cE)=r^2\,\chi(\cO_S)-\Delta(\cE)
\leq r^2\,\chi(\cO_S).
\label{Bogomolov}
\end{equation}
%%%%%%%%%%%%%%%%%%%%%%%%%%%%%%%%%%%%%%%%%%%%%%%%%%
%
%               % classical central charge

Let $J_{S}\in H_2(S;\bo{R})$ 
be a \K class on $S$.
The classical central charge  of $\cE$ measured by $J_S$
then admits an expression intrinsic to $S$:
\begin{equation}
Z_{J_S}^{\text{cl}}(\cE)
%=\bo{\varpi}_J^{\text{cl}}(f_!\,\cE)
%=-\big\langle e^{-J}\cdot v(f_!\,\cE)\big\rangle_X
=-\big\langle e^{-J_S}\cdot v_S(\cE)\big\rangle_S.
\label{BPSonS}
\end{equation}
%
%which should coincide with the exact quantum one modulo
%world sheet instanton terms.
In particular, if $J_S$ is obtained as a restriction of a \K class
$J_X$ on $X$, then (\ref{BPSonS}) coincides with
$Z_{J_X}^{\text{cl}}(f_!\,\cE)$:
the central charge measured on $X$ by $J_X$.

%%%%%%%%%%%%%%%%%%%%%%%%%%%%%%%%%%%%%%%%%%%%%%%%%%%%%%%%%
\section{String junctions versus del Pezzo surfaces}
%\\\\\\\\\\\\\\\\\\\\\\\\\\\\\\\\\\\\\\\\\\\\\\\\\\\\\\\\
\renewcommand{\theequation}{5.\arabic{equation}}\setcounter{equation}{0}
%%%%%%%%%%%%%%%%%%%%%%%%%%%%%%%%%%%%%%%%
%% %% %% %% %% %% %% %% %% %% %% %% %% %% %% %% %% %% 
%%%%%%%%%%%%%%%%%%%%%%%%%%%%%%%%%%%%%%%%%%%%%%%%%%%%%
\subsection{del Pezzo surfaces}
%%%%%%%%%%%%%%%%%%%%%%%%%%%%%%%%%%%%%%%%%%%%%%%%%%%%%
%% %% %% %% %% %% %% %% %% %% %% %% %% %% %% %% %% %%
% 
A del Pezzo surface is a surface the first Chern class of which 
is ample.
Apart from $\widetilde{B}_1\!=\!\bo{P}^1\times\bo{P}^1$, 
they are obtained by blowing up generic $N$ points 
on $\bo{P}^2$ for $0\leq N\leq 8$, 
which we call $\BN$ in this article.
Our main interest is of course in the three cases $N\!=\!8,7,6$.
The homology groups of $\BN$ are 
\begin{equation}
H_{2\bullet}\left(\BN\right)=
\bo{Z}\, [\BN]\oplus
\bo{Z}\, l\oplus \bo{Z}\, e_1\oplus\cdots \oplus \bo{Z}\, e_N
\oplus \bo{Z}\, [{\text{pt}}],
\label{homologylattice}
\end{equation}
where $l$ represents the pull-back of a line of $\bo{P}^2$, 
and $e_1,\dots , e_N$ the exceptional divisors,
by which the first Chern class is written as
$\cfBN=3l-\sum_{i=1}^Ne_i$.
We also note that the Picard group of $\BN$ is isomorphic to
$H_2(\BN)$, which means that each element of $H_2(\BN)$ is realized as
the first Chern class of a holomorphic line bundle on $\BN$ which is
unique up to isomorphism.
Intersection pairings on $H_2(\BN)$  are given by
\begin{equation}
\langle l\cdot l\rangle_{\BN}=1,\quad 
\langle l\cdot e_i\rangle_{\BN}=0,\quad
\langle e_i\cdot e_j\rangle_{\BN}=-\delta_{i,j}.
\label{intersections}
\end{equation}
We list here some topological invariants:
\begin{equation}
\big\bra \cfBN^2 \big\ket_{\BN}=9-N, \quad
\big\bra \csBN \big\ket_{\BN}= 3+N,   \quad
\chi(\cO_{\BN})=1. 
\label{topinv}
\end{equation}
The gravitational correction factor in 
the Mukai vector (\ref{gravi}) is given by
\begin{equation}
\sqrt{\frac{\Todd(T_{\BN})}{\Todd(K_{\BN})}}=
[\BN]+\frac12\,\cfBN+\frac{1}{12}(15-N)[{\text{pt}}]
\in H_{2\bullet}(\BN;\bo{Q}).
\label{gravidP}
\end{equation} 

The degree of $[C]\in H_2(\BN;\bo{Q})$ is defined by
${d}([C])=\bra [C]\cdot \cfBN\ket_{\BN}$.
If we expand it as $[C]=a_0\,l+\sum_{i=1}^Na_i\,e_i$, then 
we see  ${d}([C])=3a_0+\sum_{i=1}^Na_i$. 
It is also convenient to associate  the following two
quantities to  a \coh $\cO_{\BN}$-module $\cE$:
let $d({\cE})$ be the degree of $c_1(\cE)$, and
$k({\cE})=\bra \ch_2(\cE)\ket_{\BN}$,
in terms of which the Euler number of $\cE$ (\ref{EulercE}) can be 
expressed as
\begin{equation}
\chi(\cE) =h^0(S,\cE)-h^1(S,\cE)+h^2(S,\cE)
= r(\cE) + \frac12 d(\cE) + k(\cE).
\label{EulerdP}
\end{equation}
It is known that the degree zero sublattice of $H_2(\BN)$
is isomorphic to the $E_N$ root lattice, the root system of which 
is composed of the self-intersection $-2$ elements.  
For $3\leq N\leq 8$,  the $N$ simple roots can be chosen as
\begin{equation}
\bo{\alpha}_i=e_i-e_{i+1}, \ 1\leq i < N, \qquad 
\bo{\alpha}_N=l-e_1-e_2-e_3.
\end{equation}
%
%where from now on we  assume $N\geq 3$.
The fundamental weight 
$\bo{w}^i\in H_2(\BN;\bo{Q})$ is uniquely determined by
$\bra \bo{w}^i\cdot \bo{\alpha}_j\ket_{\BN}=-\delta_{j}^i$,
and $d({\bo{w}^i})=0$. 
Any $[C]\in H_2(\BN)$ then 
admits the following orthogonal decomposition
into the degree and  the $E_N$ weight:
\begin{equation}
[C]=\frac{{d}([C])}{9-N}\, \cfBN
+\sum_{i=1}^N\lambda_i ([C])\, \bo{w}^i,
\end{equation}
where the Dynkin labels are determined by
$\lambda_i([C])\!=\!-\bra [C]\cdot \bo{\alpha}_i\ket_{\BN}$;
for a \coh sheaf $\cE$, its Dynkin labels
$\lambda_i (\cE)$ can be defined in the same way using 
$c_1(\cE)\in H_2(\BN)$, that is,
\begin{equation}
c_1(\cE)=\frac{d(\cE)}{9-N}\,\cfBN
+\sum_{i=1}^N\lambda_i(\cE)\,\bo{w}^i.
\end{equation} 
%%%
The third intersection form 
$\chi_N:=\chi_{\BN}$ (\ref{most-natural-one})
can then be expressed as
\begin{equation}
\chi_N(\cE_1,\cE_2)
%=r_1\,r_2
%+r_1\,k(\cE_2)++r_2\,k(\cE_1)
%+\lambda(\cE_1)\cdo\lambda(\cE_2)t-\frac{1}{9-N}d(\cE_1)d(\cE_2)
%+\frac12\left(r_1d(\cE_2)-r_2d(\cE_1)\right),
=r_1\,r_2
+r_1\,k_2+r_2\,k_1
+\lambda_1\cdot\lambda_2-\frac{d_1\,d_2}{9-N}
+\frac12\,\left(r_1\,d_2-r_2\,d_1\right).
\label{kaiN}
\end{equation} 
For more information on \dP surfaces see, for example,
\cite{MNW,DHIK,HI,MS,DKV,FMW}.

$\BN$ has  a natural one-parameter family of {\it complexified}
\K classes $J_S=t\,\cfBN$, with ${\text{Im}}(t)\!>\!0$,
because  $\cfBN$ is  an ample divisor.
The degree of a curve defined above is nothing but the volume of
it measured by the normalized \K class $J_S=\cfBN$.

Using (\ref{RRonS}) combined with (\ref{gravidP}),
it is now straightforward to compute the classical central charge
of a D-brane $\cE$ measured by the \K class $J_S=t\,\cfBN$:
\begin{align}
Z_{t}^{\text{cl}}(\cE)
&=-\big\bra e^{-t\cfBN}v_{\BN}(\cE)\big\ket_{\BN} 
\nonumber \\
&=-r(\cE)\,(9-N)\left(\frac{t^2}{2}-\frac{t}{2}
+\frac{1}{12}\,\frac{3-N}{9-N}\right)
+d(\cE)\, t- \chi(\cE).
%\left(r+k(\cE)+\frac12 d(\cE) \right).
\label{BPSmass}
\end{align}
%
%\\\\\\\  quantum charge   \\\\\\\\\\\\\\\\\\\\\\\\\\\\\\\\\
Recall that the exact quantum central charge 
$Z(\cE)$ yields the classical one evaluated above (\ref{BPSmass})
modulo the instanton correction terms  
$O(e^{2\pi i t})$.
In particular, for the cases $N=8,7,6$, the instanton expansion
of the central charge (\ref{eqn:lr}) of the D4-brane obtained in the 
previous section takes the form:
\begin{equation}
t_d=\left(\frac{t^2}{2}-\frac{t}{2}
+\frac{1}{12}\,\frac{3-N}{9-N}\right)+O(e^{2\pi i t}),
\end{equation}
in writing which we have noticed that $4\sin^2\pi\alpha=9-N$ 
for $\widehat{E}_{N=8,7,6}$.
The classical part of $t_d$ coincides with the one in (\ref{BPSmass}).
Therefore  we can make the following identification of
the quantum central charge of the D-brane $\cE$ measured by the
\K form $tc_1(\BN)$  by
\begin{equation}
Z_{t}(\cE)=-r(\cE)\,(9-N)\,t_d+d(\cE)\,t-\chi(\cE).
\label{chusin}
\end{equation}
%%%%%
We are now in a position to compare the two 
central charges of the {\em one and the same} $\EN$ theory; %for $N=8,7,6$;
one obtained by the analysis of the SW periods (\ref{eqn:cc}), 
and the other by the geometric method (\ref{chusin}).
They are related under mirror symmetry through (\ref{eqn:cc2})
so that we have the following dictionary between the charges, after 
a trivial rescaling of the former:
\begin{align}
p&=d(\cE), \nonumber\\
q&=r(\cE),         \label{qr-taiou}\\
-n&=\chi(\cE)=r(\cE)+\frac12 d(\cE)+k(\cE).\nonumber
\end{align} 
%

%
%the \K modulus dependence of which is characterized by the \PF equation:
%\begin{equation}
%\big\{\vartheta^2-(\vartheta+\alpha)(\vartheta+1\!-\!\alpha)\big\}\,
%\vartheta\, Z(\cE)=0,
%\end{equation}
%where $\alpha=1/3,1/4, 1/6\ $  for $N=6,7, 8,\ $  respectively.
%\\\\\\\\\\\\\\\\\\\\\\\\\\\\\\\\\\\\\\\\\\\\\\\\\

In passing, we give a comment on the $N\!=\!0$ case. 
Upon a change of variables:
$t=\frac13t_b+\frac12$, where the constant shift implies 
the existence of the NS B-field flux on $\bP^2$ \cite{DG},
(\ref{BPSmass}) becomes
\begin{equation}
Z^{\text{cl}}(\cE)=
-\left\bra e^{-t_b l}\ch(\cE)\sqrt{\frac{\widehat{A}(T_{\bo{P}^2})}
{\widehat{A}(N_{\bo{P}^2})}}
  \right\ket_{\bo{P}^2}
%t=\frac13t_b+\frac12, D=\frac13 c_1(\bo{P}^2),\\
=-r(\cE)\,\left(\frac{t_b^2}{2}+\frac18 \right)
+\frac13 d(\cE)\, t_b-k(\cE),
\end{equation}
where $l$ is the ample generator of divisors.
This is precisely the classical central charge on $\bo{P}^2$ 
treated as the $\bo{Z}_3$  orbifold in \cite{DG,DFR,MOY}.
%\\\\\\\\\\\\\\\\\\\\\\\\\\\\\\\\\\\
%
%% %% %% %% %% %% %% %% %% %% %% %% %% %% 
\subsection{$\bo{E}_{\bf 9}$ almost del Pezzo surface}
%% %% %% %% %% %% %% %% %% %% %% %% %% %% 
An $E_9$ almost del Pezzo surface $\Bnine$ is a surface obtained 
by blowing up nine points of $\bo{P}^2$ which are the complete 
intersection of two cubics on $\bo{P}^2$.
It has the  structure of elliptic fibration
%\begin{equation}
$\pi\colon\Bnine\to \bo{P}^1$,
%\label{elliptic}
%\end{equation} 
%
which has twelve degenerate fibers leaving the total space non-singular
for a generic choice of parameters, which we assume throughout the paper.
As its name stands for, $\Bnine$ shares many properties
with the del Pezzo surfaces $\BN$;
to be explicit, among the formulae in the preceding subsection,
%-------------
(\ref{homologylattice})--(\ref{EulerdP})
%---------------
remain valid for $\Bnine$ if one simply sets $N\!=\!9$ there,
as well as the definition of the first Chern class
$c_1(\Bnine)$.       
% and that of the degree of a curve.      
It is then clear that the elements of $H_2(\Bnine)$ 
orthogonal to both $c_1(\Bnine)$ and $e_9$ 
generate the $E_8$ root lattice. 
Let $[{\text{F}}]$ and $[{\text{B}}]$ be the class of $H_2(\Bnine)$ 
defined by the fiber and a cross section of the fibration respectively. 
We can then make the following identification:
\begin{equation}
[{\text{F}}]=c_1(\Bnine)=3l-\sum_{i=1}^9e_i,\quad
[{\text{B}}]=e_9,
\end{equation}
the intersection pairings of which we give here for convenience:
\begin{equation}
\big\bra [{\text{F}}]\cdot [{\text{F}}]\big\ket_{\Bnine}=0,\quad
\big\bra [{\text{F}}]\cdot [{\text{B}}]\big\ket_{\Bnine}=1,\quad
\big\bra [{\text{B}}]\cdot [{\text{B}}]\big\ket_{\Bnine}=-1.
\end{equation}
As opposed to the case of del Pezzo surfaces, 
$c_1(\Bnine)$ is no longer an ample divisor;
it is only a nef divisor with self-intersection zero,
so that $c_1(\Bnine)$ alone cannot define a \K class on $\Bnine$.
However, the structure of elliptic fibration  suggests the following 
natural two-parameter family of the complexified \K classes on $\Bnine$:
\begin{equation}
J=t_1[{\text{F}}]+t_2\left([{\text{F}}]+[{\text{B}}]\right)
=(t_1+t_2)\,c_1(\Bnine)+t_2\,e_9,
\label{two-parameter}
\end{equation}  
where the imaginary parts of $t_1$ and $t_2$
parametrize the volume of the curve 
$[{\text{B}}]$ and $[{\text{F}}]$ measured by ${\text{Im}}\,(J)$ respectively,
which span the \K sub-cone of our model. 
%%%%%%%%%%%%%%%
A serious treatment of the stability condition of 
\coh $\cO_{\Bnine}$-modules would face with 
the problem of subdivision of the \K cone, because
the stability condition depends on the choice of 
\K class ${\text{Im}}\,(J)$ \cite{ShK}, which we will not discuss further
in this article. 
%We also note here that 
By the way, another two-parameter family of \K classes 
$\tilde{J}$ treated in \cite{HST} can be written in our notation as  
$\tilde{J}=\tilde{t}_1\,[{\text{F}}]+\tilde{t}_2\,l$. 
  
%%%%%%%%%%%%%%%
The classical central charge of
the D-brane represented by  a \coh $\cO_{\Bnine}$-module
$\cE$ can be computed as:
\begin{align}
Z^{\text{cl}}(\cE)&=-\left\bra
e^{-t_1[{\text{F}}]-t_2([{\text{F}}]+[{\text{B}}])}\cdot
\ch(\cE)\cdot\big([\Bnine]+\frac12[{\text{F}}]+\frac12 [{\text{pt}}]\big)  
\right\ket_{\Bnine}\nonumber \\
&=-r(\cE)\left(\frac{t_2^2}{2}+t_1t_2-\frac{t_2}{2}-\frac12\right)
+d(\cE)\,(t_1+t_2)+b(\cE)\,t_2
-\chi(\cE),
\label{centerB9}
\end{align}
where $d(\cE)$ and $b(\cE)$ are the two integers defined by the  
decomposition of $c_1(\cE)$: 
\begin{equation}
c_1(\cE)=b(\cE)\,[{\text{F}}]+d(\cE)\,([{\text{B}}]+[{\text{F}}])
+\sum_{i=1}^8 \lambda_i(\cE)\, \bo{w}^i.
\end{equation}
That is, they are obtained  via 
%\begin{equation*}
$b(\cE)=\bra [{\text{B}}]\cdot c_1(\cE)\ket_{\Bnine}$,
$d(\cE)=\bra  [{\text{F}}]\cdot c_1(\cE)\ket_{\Bnine}$.
%\end{equation*}
In terms of these variables, the intersection form 
$\chi_9:=\chi_{\Bnine}$ can be written as 
\begin{equation}
\chi_9(\cE_1,\cE_2)=r_1\,r_2
+r_1\,k_2+r_2\,k_1 
-d_1\,d_2-b_1\,d_2-b_2\,d_1
+\lambda_1\cdot\lambda_2
+\frac12(r_1\,d_2-r_2\,d_1).
\label{kai9}
\end{equation}
%

%%%%%%%%%%%%%%
%\vspace{2cm}
%%%%%%%%%%%%%%
%
A period $\varpi$ obeys the Picard--Fuchs differential equations:
%\begin{equation}
$ 
{\cal L}^{(1)}\varpi=0, \
{\cal L}^{(2)}\varpi=0,
%\end{equation}
$
with 
\begin{align}
{\cal L}^{(1)}&=\big(\vartheta_1
-z_1\left(\vartheta_1-\vartheta_2\right)\big)\,
\vartheta_1, 
\label{B9P-F1}\\
%  -  -  -  -  -  -
{\cal L}^{(2)}&=\vartheta_2\left(\vartheta_2-\vartheta_1\right)
-z_2(\vartheta_2+\tfrac16)(\vartheta_2+\tfrac56),
\label{B9P-F2} 
\end{align}
which can be obtained from the standard procedure of
the local mirror principle \cite{CKYZ} using 
a realization of $\Bnine$ as a hypersurface in a toric threefold.     
%the {\em bare} \K moduli parameters as usual.
It is easy to see that we can take
the two periods of the $\widehat{E}_8$ torus:
$$
\varpi_0(z_2)={}_2F_1\big(\tfrac16,\tfrac56;1;z_2\big), \quad 
\varpi^{(2)}_1(z_2)
=\frac{1}{2\pi i}\,\big(\varpi_0(z_2)\log(\frac{z_2}{432})+
%\frac{1}{(2\pi i)}\,
{{}_2F_1}^*(\tfrac16,\tfrac56;1;z_2)\big),
$$ 
as two of the four periods of the \PF system (\ref{B9P-F1}), 
(\ref{B9P-F2})
with $t_i=\varpi^{(i)}_1/\varpi_0$
$(i=1,2)$; the mirror map. 
As for the D4-brane period $\varpi_2$, 
we can take it to have the following form at the large radius limit:
\begin{equation}
t_d:=\frac{\varpi_2}{\varpi_0}=
\left(\frac{t_2^2}{2}+t_1t_2-\frac{t_2}{2}-\frac12\right)
+O(e^{2\pi i t_1},e^{2\pi i t_2}),
\label{ansatz-D4}
\end{equation}
%
%according to the classical limit 
%of the central charge (\ref{centerB9}).
because this is the classical part of 
the only four-cycle period, modulo addition of
periods of lower dimensional cycles, that remains finite 
under the limit of infinite elliptic fiber 
of a \cy threefold  which contains
$\Bnine$ as a section of elliptic fibration \cite{CKYZ}.
Moreover, rewritten in the new variables 
${\cal U}=t_2$, $\tilde{\phi}=t_1+t_2$, 
(\ref{ansatz-D4}) coincides with the period  
$\tilde{\phi}_D$ of the Phase I in \cite{LMW}.
Thus in terms of the basis of the solutions of the \PF equations:
$\{ \varpi_0,\varpi^{(1)}_1,\varpi^{(2)}_1,\varpi_2 \}$,
the quantum central charge of the \coh sheaf $\cE$ on $\Bnine$ 
measured by $J$ (\ref{two-parameter}) can be expressed up to normalization
factor as 
\begin{align}
Z(\cE)&=\frac{1}{\varpi_0}\left(
-r(\cE)\,\varpi_2
+d(\cE)\,(\varpi^{(1)}_1+\varpi^{(2)}_1)
+b(\cE)\,\varpi^{(2)}_1
-\chi(\cE)\,\varpi_0\right),\nonumber \\
&=-r(\cE)\,t_d+d(\cE)\,(t_1+t_2)+b(\cE)\,t_2-\chi(\cE).
\label{B9center}
\end{align}
We leave further detailed investigation  of this \PF system to a future work.  
%
%%%%%%%%%%%%%%%%%%%%%%%%%%%%%%%%%%%%%%%%%%%%%%%%%%%%%%%%%%%%%%%%%%%%%
%%%%%%%%%%%%%%%%%%%%%%%%%%%%%%%%%%%%%%%%%%%%%%%%%%%%%%%%%%%%%%%%%%%%%
%\newpage
\subsection{Duality maps}
\label{sec:map}
For the $\bo{\hat{E}}_N$ 7-brane configuration 
in the type IIB side, where we restrict ourselves to the cases 
$3\leq N\leq 8$ for simplicity, let us recapitulate 
a string junction $\bo{J}$ as given in (\ref{eqn:junction2}):
\begin{equation}
\bo{J}=\sum_{i=1}^N\lambda_i\,\om^i+q\,\om^q
+p\,\om^p+n\,\bo{\delta}^{(-1,0)},
\end{equation}
where $(\lambda_i)$ is the $E_N$ weight vector, 
$(p,q)$ the asymptotic charge, and $n$ the grade of the junction.
The following intersection form $\varPhi_N$ on the junction lattice
is adopted in \cite{HI}: 
\begin{equation}
\varPhi_N(\bo{J}_1,\bo{J}_2)=
-\lambda_1\cdot\lambda_2+n_1q_2+n_2q_1+\frac{p_1p_2}{9-N}
+q_1q_2+ p_1q_2.
\end{equation}
%
%%%
As discussed in sections 2 and 3, 
we have two realizations of  BPS states in 
the $E_N$ theories on $\bo{R}^4\!\times\!S^1$:
either by a \coh $\cO_{\BN}$-module $\cE$ in the type IIA side or
by a string junction $\bo{J}$ in the type IIB side, 
which raises a natural question:
what is the correspondence between coherent sheaves on a del Pezzo surface
$\BN$ and string junctions in the $\bo{\hat{E}}_N$ 7-brane background?
An answer has been given by Hauer and Iqbal \cite{HI}, 
who found that the following map%
%-----------------------
\footnote[2]{Strictly speaking,
$\bo{J}(\cE)$ as well as  $Z(\cE)$ depend only on 
$\ch(\cE)$, but the notation adopted here will cause no confusions.}
%-----------------------
%
\begin{equation}
\ch(\cE) \rightarrow
 \bo{J}(\cE)=\sum_{i=1}^N \lambda_i (\cE)\,\om^i
+r(\cE)\,\om^q+d(\cE)\,\om^p-\chi(\cE)\,\bo{\delta}^{(-1,0)}
\label{EJcor}
\end{equation}
induces an isomorphism of the junction lattice and 
the homology lattice $H_{2\bullet}(\BN)$, which we identify with
the RR charge lattice of D-branes on $\BN$, that is,
\begin{equation}
\varPhi_N(\bo{J}(\cE_1),\bo{J}(\cE_2))=-\chi_N(\cE_1,\cE_2).
\label{isom}
\end{equation}
The inequality constraint (\ref{Bogomolov})
imposed on the self-intersection of a 
torsion-free  semi-stable sheaf $\cE$ corresponding to a BPS brane 
is then converted  into  
$\varPhi_N(\bo{J},\bo{J})\geq -q^2$,
which is surely a {\em necessary} condition for a junction $\bo{J}(\cE)$
to represent a BPS state.
This may serve as a physical consistency check
of the map proposed above (\ref{EJcor}).
%

%
%%%%%%%%%%%%%%%%%%%%%%%%%%
The map (\ref{EJcor}) has been obtained by the 
inspection of the two intersection forms:
$\varPhi_N$ on the $\EN$ junction lattice
and $\chi_{N}$ on the homology lattice $H_{2\bullet}(\BN)$.
On the other hand, we have identified the $\EN$ string junction charges 
with the central charges of D-branes on $\BN$ in (\ref{qr-taiou}),
which leads us to define a natural map $\rho_N$ from 
the string junctions (\ref{EJcor})
to the D-brane central charges on the del Pezzo surface (\ref{chusin})
measured by the \K class $t\cfBN$ by 
\begin{equation}
\rho_N(\om^q,\om^p,\om^i,\bo{\delta}^{(-1,0)})
= (-(9-N)\,t_d,t,0,1), 
\end{equation}
so that we have the {\em correspondence}:
\begin{equation}
\rho_N(\bo{J}(\cE))=Z_t(\cE), 
\label{ZJisomN}
\end{equation}
which we propose as another evidence for the map (\ref{EJcor}).
Note that the junctions carrying $E_N$ weights 
$(\om^i)$ are in the kernel of $\rho_N$, only because
our \K class $t\cfBN$ cannot see $E_N$ quantum numbers;
it should not be so difficult to incorporate $E_N$ quantum numbers
in the central charge $Z(\cE)$, see \cite{MNW,MNVW}.

%\\\\\\\\\\\\\\\\\\\\\\\\\\\\\\\\\\\\\\\\\\\\\\\\\\\\\\\\\\\\\
Now that the correspondence between 
\coh $\cO_{\BN}$-modules and $\EN$ string junctions 
has been found, our next task is to explicitly construct 
the string junctions for various \coh $\cO_{\BN}$-modules.  
We give here the images of the map (\ref{EJcor})
of a few basic \coh sheaves on $\BN$:
\begin{align}
\bo{J}(\cO_{\BN})&=\om^q-\bo{\delta}^{(-1,0)},  \label{Dfour}\\
\bo{J}(\iota_!\, L_C)&=
\sum_{i=1}^N\lambda_i([C])\,\om^i   +   d([C])\,\om^p 
-\chi(L_C)\,\bo{\delta}^{(-1,0)},     \label{Dtwo}\\
\bo{J}(\cO_p)&=-\bo{\delta}^{(-1,0)},           \label{Dzero}
\end{align}
where (\ref{Dfour}) is a D4-brane wrapped on $\BN$; 
$\iota\colon C\!\hookrightarrow\!\BN$ in (\ref{Dtwo})
is an embedding of curve, 
and $L_C$ an line bundle on $C$, which represents a D2-brane wrapped on
$C$ bounded with several D0-branes;
finally $\cO_p$ in (\ref{Dzero}) is called 
the {\em skyscraper} of length one with support
at a point $p\in \BN$, which clearly corresponds to a D0-brane at $p$.
%
%Note also that among three sheaves above, 
%the first is rigid, while the latter two has moduli.
%

To be more explicit in D2-branes (\ref{Dtwo}), 
we want to describe two typical examples here:
For the first example, let $C$ be an {\em exceptional curve}, that is,
a rational curve with self-intersection $-1$, 
so that $d([C])=1$ by the classical adjunction formula
(\ref{adjunction}).
The totality of the exceptional curves
spans the fundamental Weyl orbit of $E_N$:
$({\bf 3},{\bf 2})$, ${\bf 10}$,
${\bf 16}$, ${\bf 27}$, ${\bf 56}$, ${\bf 240}$, for $3\leq N \leq 8$. 
Next we take $\cO_C$ as a line bundle on it;
then the corresponding $\EN$ string junction becomes
\begin{equation}
\bo{J}(\iota_!\,\cO_C)=\sum_{i=1}^N\bar{\lambda}_i\,\om^i
+\om^p - \bo{\delta}^{(-1,0)},
\quad \bar{\lambda}\cdot \bar{\lambda}=\frac{10-N}{9-N}.
\end{equation}
%
%where $(\bar{\lambda}_i)$ above represents one of the weights 
%of the fundamental representation of $E_N$.
%
For the second example, we take an elliptic curve $E$ with
$[E]=\cfBN\in H_2(\BN)$, which is known as 
an {\em anti-canonical divisor} in $\BN$; 
we consider also a degree zero line bundle $L_E$ on it,
which is parametrized by the Jacobian of $E$: ${\text{Jac}}(E)\cong E$;
it is easy to see that the corresponding string junction becomes
\begin{equation}
\bo{J}(\iota_!\,L_E)=(9-N)\,\om^p.
\end{equation}
%
%Note that this junction has neither $E_N$- nor loop junction charge.
%\\\\\\\\\\\\\\\\\\\\\\\\\\\\\\\\\\\\\\\\\\\\\\\\\\\\\\\\\\\\\\\\\\\

%\\\\\\\\\\    E_9  Almost del Pezzo    \\\\\\\\\\\\\\\\\\\\\\\\\\\
It is possible to extend the above considerations to
the $\widehat{E}_9$ theory.
The string junction  and intersection form in this case \cite{HI}
are given by
\begin{align}
\bo{J}&=\sum_{i=1}^8\lambda_i\,\om^i+p\,\om^p+q\,\om^q+
n\,\bo{\delta}^{(-1,0)}+m\,\bo{\delta}^{(0,1)},\\
\varPhi_9(\bo{J}_1,\bo{J}_2)&=
-\lambda_1\cdot\lambda_2
+p_1p_2+p_1q_2+q_1q_2
+m_1p_2+p_1m_2
+q_1n_2+n_1q_2.
\end{align}
Hauer and Iqbal have shown that the following map from
the homology lattice $H_{2\bullet}(\Bnine)$ in the type IIA side to
the $\widehat{E}_9$ string junctions in the type IIB side:
\begin{equation}
\ch(\cE)\to \bo{J}(\cE)=
\sum_{i=1}^8\lambda_i(\cE)\,\om^i+d(\cE)\,\om^p+r(\cE)\,\om^q
+b(\cE)\,\bo{\delta}^{(0,1)}-\chi(\cE)\,\bo{\delta}^{(-1,0)},
\label{EJ2}
\end{equation}
again defines the isomorphism of the lattices:
\begin{equation}
\varPhi_9(\bo{J}(\cE_1),\bo{J}(\cE_2))=-\chi_9(\cE_1,\cE_2).
\label{isom2}
\end{equation}
According to our point of view, on the other hand, 
comparison of (\ref{EJ2}) with (\ref{B9center})
leads to define a natural  map $\rho_9$ from 
the $\widehat{E}_9$ string junctions 
to the $\Bnine$ central charges by
\begin{equation}
\rho_9 \left(\om^q,\om^p,\om^i,\bo{\delta}^{(0,1)},
\bo{\delta}^{(-1,0)}\right)
=\left(-t_d,t_1+t_2,0,t_2,1\right),
\end{equation} 
which again induces the {\em correspondence}
\begin{equation}
\rho_9(\bo{J}(\cE))=Z_{t_1,t_2}(\cE).
\label{ZJisom9}
\end{equation}
%
%%%%%%%%%%%%%%%

To sum up, what we have done  can be 
succinctly shown in the commutative diagram:
%\vspace{0.5cm}
%\\\\\\\\\\\\\\\
\begin{center}
\setlength{\unitlength}{1cm}
\begin{picture}(5,2.8)
%\put(0,0){\framebox(4.5,2.3)}
\put(1.8,2.3){$\cong$}
\put(1.5,0.2){$Z_t(\cE)$}
\put(0,2){$\ch(\cE)$}
\put(3,2){$\bo{J}(\cE)$}
\put(3,1.2){$\rho_N$}
\put(1.6,2.1){\vector(1,0){0.8}}
\put(0.9,1.6){\vector(1,-1){0.7}}
\put(2.9,1.6){\vector(-1,-1){0.7}}
\end{picture}
\end{center}
%\\\\\\\\\\\\\\
where the horizontal arrow is the isomorphisms 
(\ref{EJcor}), (\ref{EJ2}) proposed in \cite{HI},
while the remaining two are ours; the left being the central charge formulae
(\ref{chusin}), (\ref{B9center}), and the right the correspondences
(\ref{ZJisomN}), (\ref{ZJisom9}).

%\\\\\\\\\\\\\\\\\\\\\\\\\\\\\\\\\\\\\\\\\\\\\\\\\\\\\\\ 
\section{Conclusions}

In this article we have first derived the central charge formula for the
D3-brane probe theory in the $\bo{\hat{E}}_{N=8,7,6}$ 7-brane 
backgrounds. Employing local mirror symmetry we then translate the central
charge for the IIB string junctions into that for IIA D-branes on del Pezzo
surfaces $B_{N=8,7,6}$. To make this precise we have compared the quantum
central charge (modulo instanton corrections) with the classical central
charge verified by the geometric analysis of D-brane configurations.
As a result, we have shown that the duality maps (\ref{EJcor}), (\ref{EJ2})
between the homology lattice of the del Pezzo
surface $H_{2\bullet}(\BN)$
and the $\EN$ string junction lattice,
originally found in \cite{HI} based 
on the isomorphism of the lattices $\varPhi_N$ (\ref{isom}), (\ref{isom2}),
can be naturally recovered from the correspondence of
the string junctions and the del Pezzo central charges 
(\ref{ZJisomN}), (\ref{ZJisom9}) for $N=9,8,7,6$ and the $E_N$ singlet part.
The cases for $N\leq 5$ are also worth being examined in detail, and
will be analyzed in our subsequent paper \cite{MOY2}.
Since the $E_{N\leq 4}$ theories on $\bo{R}^4\times S^1$ reduce
to 4D asymptotically free theories in the $R\rightarrow 0$ limit
it will be interesting to compare with a recent work \cite{EK}
in which $SU(2)$ gauge theory with fundamental matters on $\bo{R}^4\times S^1$
is investigated by compactifying the type II theory on the local
$\bo{F}_2$.

For further directions in future study, let us 
note that the duality maps (\ref{EJcor}), (\ref{EJ2})
concern only with the RR charges on the del Pezzo side or with
the junction charges on the 7-brane side, 
while an actual \coh $\cO_{\BN}$-module $\cE$ or an $\EN$
string junction have moduli parameters in general;
for example, on the del Pezzo side, the structure sheaf 
$\cO_{\BN}$ is rigid, 
while a torsion sheaf  $\iota_!\,L_E$
with support on an anti-canonical divisor $E$ 
clearly has moduli parameters.
Therefore it would be quite interesting to establish the duality map
between \coh $\cO_{\BN}$-modules and $\EN$ string junctions
{\em including their moduli parameters}.

Another important issue is to analyze the stability of D-branes on del Pezzo
surfaces. The stability of BPS-branes is the subject of current interest
\cite{HM,Gepner,ref-stab}. Under the map (5.25) it will be possible to study the
stability of certain D-brane configuration on $B_N$ in terms of the
corresponding junction configuration in the $\bo{\hat{E}}_N$ 7-brane 
background. Then the most interesting task is to determine curves of marginal
stability (CMS) of BPS states and follow their decay processes. CMS of BPS
junctions may be worked out numerically. Our preliminary computation indicates
that there appear infinitely many CMS on the $u$-plane and their patterns look
quite different from the ones observed for ordinary 4D ${\cal N}\!=\!2$ 
$SU(2)$ gauge theories. We hope to report the results elsewhere 
in the near future.

\vskip6mm\noindent
{\bf Acknowledgements}

\vskip2mm
S.K.Y. would like to thank Y. Yamada for interesting discussions.
The research of Y.O. is supported by JSPS Research Fellowship for Young 
Scientists.
The research of K.M. and S.K.Y. was supported in part by Grant-in-Aid for 
Scientific Research 
on Priority Area 707 ``Supersymmetry and Unified Theory of Elementary 
Particles'', Japan Ministry of Education, Science and Culture.

\newpage

%%%%%%%%  References          
%%% Please include the hep-th number at the end.

%%%%%%%%%%%%%%%%%%%%%%%%%%%%%%%%%%%%%
\end{document}